\documentclass[12pt]{iopart}

\usepackage{graphicx,subfigure,iopams}

\begin{document}
\article[A new treatment of nonlocality in scattering process]{Article}
{A new treatment of nonlocality in scattering process}

\author{N. J. Upadhyay$^{\dagger}$, A. Bhagwat$^{\dagger\dagger}$ and
B. K. Jain$^*$}
\address{UM-DAE Centre for Excellence in Basic Sciences, Vidyanagari,
Mumbai-400098, India}
\ead{$^{\dagger}$neelam.upadhyay@cbs.ac.in, $^{\dagger\dagger}$ameeya@cbs.ac.in,
$^*$brajeshk@gmail.com}

\begin{abstract}
Nonlocality in the scattering potential leads to an integro-differential
equation. In this equation nonlocality enters through an integral
over the nonlocal potential kernel. The resulting Schr\"odinger equation
is usually handled by approximating $r$,$r^\prime$-dependence of the
nonlocal kernel. The present work proposes a novel method to solve the
integro-differential equation. The method, using the mean value theorem
of integral calculus, converts the nonhomogeneous term to a homogeneous
term. The effective local potential in this equation turns out to be
energy independent, but has relative angular momentum dependence. This
method has high accuracy and is valid for any form of nonlocality. As
illustrative examples, the total and differential cross sections for
neutron scattering off $^{12}$C, $^{56}$Fe and $^{100}$Mo nuclei are
calculated with this method in the low energy region (up to 10 MeV)
and are found to be in reasonable accord with the experiments.
\end{abstract}
\maketitle

\section{Introduction}
The nucleon-nucleus interaction is known to be nonlocal in 
nature \cite{bethe,frahn1,frahn2}. This nonlocal character arises
because of the many-body effects such as the virtual excitations in
the nucleus and the exchange of the nucleons within the interacting
system \cite{frahn1,frahn2,lemere,balan97}. The explicit use of the nonlocal
interaction framework, therefore, enriches the theoretical description
of the scattering process. Its incorporation, however, leads to the
integro-differential form of the Schr\"{o}dinger equation, which is
difficult to solve. It is written as:
\begin{equation}
\left[\frac{\hbar^2}{2\mu}\nabla^2 + E + V_{SO}{\bf L}\cdot{\boldsymbol \sigma}
\right]\Psi({\bf r}) = \int V({\bf r},{\bf r^\prime})\Psi({\bf r^\prime})
d{\bf r^\prime}
\label{eq1}
\end{equation}
where $\mu$ is the reduced mass of the nucleon-nucleus system, $E$
is the center of mass energy, ($V_{SO}\,{\bf L}\cdot{\boldsymbol \sigma}$) is
the local spin-orbit interaction, $V({\bf r},{\bf r^\prime})$ is the
nonlocal interaction kernel and $\Psi({\bf r})$ is the scattering 
wave function.

Any effort to solve Eq.(\ref{eq1}) requires the explicit form of
$V({\bf r},{\bf r^\prime})$, which, unfortunately, is not known. However,
it is expected that this form should be such that in the limit of
vanishing nonlocality, the integro-differential equation reduces to
the conventional homogeneous equation. Guided by this idea, a factorized 
form for the nonlocal kernel was proposed by Frahn and Lemmer
\cite{frahn1,frahn2}, which is written as
\begin{equation}
V({\bf r},{\bf r^\prime})= \frac{1}{\pi^{3/2}\beta^3}\,
{\rm exp}\left[-\,\frac{\left|{\bf r} - {\bf r^\prime}\right|^2}{\beta^2}
\right] U \left(\frac{\left|{\bf r} + {\bf r^\prime}\right|}{2}\right)
\label{eq2}
\end{equation}
where $\beta$ is the range parameter and $U$ is the nonlocal interaction.
The term with $\beta$ represents the behaviour of nonlocality. It reduces
to the Dirac $\delta$ function in the limit of vanishing nonlocality. 
This prescription has been used by several groups, some of the notable
amongst them being the work of Perey and Buck \cite{pb} and Tian {\it et al.} 
\cite{tpm15}. The important aspect of the former study \cite{pb} is the
construction of the energy dependent local equivalent potential which
can be used in the homogeneous Schr\"{o}dinger equation. This result is
obtained by using the gradient approximation, which is found to be reliable
for tightly bound nuclei. However, for nuclei away from the stability line 
its validity might be a suspect. In the other study, Tian {\it et al.}
\cite{tpm15} have treated nonlocality by solving Eq.(\ref{eq1}) using
Lanczos method \cite{kimUd1,kimUd2}.

Other approaches to obtain efficient solution to Eq.(\ref{eq1}) with a 
general nonlocal kernel include writing the nonlocal kernel in separable
form \cite{ali,ahmad}, expanding nonlocal kernel in terms of Chebyshev
polynomials \cite{raw3}, to name a few. Recently a microscopic approach 
to address nonlocality has been developed by Rotureau {\it et al.} 
\cite{rot17} wherein the nonlocal optical potential for nucleon-nucleus
scattering is constructed from chiral interactions. These methods, 
however, might have computational limitations in analyzing the 
nucleon-nucleus scattering data routinely.

In this article we present a readily implementable technique to solve
the integro-differential Schr\"{o}dinger equation. With a very simple
approximation, this technique reduces the integro-differential equation
to a homogeneous differential equation. This is achieved by using the
mean value theorem (MVT) of integral calculus \cite{mvt}. Application
of the MVT converts the nonlocal interaction to a local form. This 
local potential is energy independent, but depends upon relative
angular momentum ($l$). The important aspect of this method is that
it does not depend upon any particular choice of the nonlocal form
factor and computationally it is very tractable.

The paper is organized as follows: In Section 2 we present the
MVT technique used to reduce the integro-differential equation to a
homogeneous one and identify the requirements for its applicability. 
The accuracy of the solution of the homogeneous equation is established
in Section 3. As our method is applicable to any choice of the 
nonlocal form factor, we study its applicability for different choices
in Section 4. Finally, Section 5 is devoted to the comparison of the 
predictions of our method with the experimental observables like, total 
and differential cross sections. As illustrations, we have studied
neutrons scattering off $^{12}$C, $^{56}$Fe and $^{100}$Mo targets.

\section{Method to solve the integro-differential equation}
For the present, dropping the spin-orbit term in Eq.(\ref{eq1}) we write
partial wave expansion for the scattering wave function, $\Psi$, and
the nonlocal potential, $V({\bf r},{\bf r^\prime})$, as
\begin{eqnarray}
&&\Psi({\bf r})\,=\,\sum_{l,m_l} i^l \,\frac{u(l;r)}{r}\,Y_{lm_l}(\Omega_r)
\,\,\,\,\,{\rm and}
\label{eq3}
\\
&&V({\bf r},{\bf r^\prime})\,=\,\sum_{l^\prime,m^\prime_l} 
\frac{g(l^\prime;r,r^\prime)}{rr^\prime} \,Y_{l^\prime m^\prime_l}(\Omega_r)\,
Y^*_{l^\prime m^\prime_l}(\Omega_{r^\prime})
\label{eq4}
\end{eqnarray}
respectively. The resulting Schr\"{o}dinger equation for the 
$l^{\rm th}$ partial wave is 
\begin{equation}
\left[\frac{d^2}{dr^2} - \frac{l(l+1)}{r^2} + 
\frac{2\mu\,E}{\hbar^2} \right] u(l;r)\,=\,
\frac{2\mu}{\hbar^2}\int_0^{r_m} g(l;r,r^\prime) 
u(l;r^\prime)\,dr^\prime\,.
\label{eq5}
\end{equation}
The upper limit of the integration over nonlocal kernel, 
$\displaystyle{g(l;r,r^\prime)}$, is the matching radius ($r_m$)
at which its contribution to the integral becomes negligible. 
The nonlocal kernel for $l^{\rm th}$ partial wave is written as
\begin{eqnarray}
&& g(l;r,r^\prime)=\left(\frac{2r r^\prime}{\sqrt{\pi}\,\beta^3}\right)
{\rm exp}\left(\frac{-r^2 - {r^\prime}^2}{\beta^2}\right)
\label{eq6}
\\
\nonumber
&&\hspace{0.5cm}\times
\int_{-1}^{1} U \left(\frac{|{\bf r}+{\bf r^\prime}|}{2}\right) 
{\rm exp}\left(\frac{2rr^\prime {\rm cos}\,\theta}{\beta^2}\right) 
P_l\left({\rm cos}\,\theta\right) d\left({\rm cos}\,\theta\right),
\end{eqnarray}
where $\theta$ is the angle between the vectors ${\bf r}$ and 
${\bf r^\prime}$ \cite{pb}. In the literature \cite{pb}, $|{\bf r}
+{\bf r^\prime}|$ is approximated as $(r+r^\prime)$ leading to
\begin{eqnarray}
g(l;r,r^\prime)&=&\left(\frac{2r r^\prime}{\sqrt{\pi}\,\beta^3}\right)
{\rm exp}\left(\frac{-r^2 - {r^\prime}^2}{\beta^2}\right)
U \left(\frac{r+r^\prime}{2}\right)
\label{eq7}
\\
\nonumber
&&\hspace{1.5cm}\times
\int_{-1}^{1} {\rm exp}\left(\frac{2rr^\prime {\rm cos}\,\theta}
{\beta^2}\right) P_l\left({\rm cos}\,\theta\right) d\left({\rm cos}\,
\theta\right)\,.
\end{eqnarray}
However, in the present work we do not use the above approximation. The
integral appearing in Eq.(\ref{eq6}) is evaluated numerically. Since
Eq.(\ref{eq7}) is used very often in the literature, effect of the
approximation leading to it is examined explicitly later in Section 3.2. 

\subsection{The nonlocal potential, $U(r)$}
The construction of the non-local kernel requires the nucleon-nucleus
potential, $U(r)$. Since in the present work our aim is to propose a new 
treatment of nonlocality and establish its correctness, we prefer to have
a well-established prescription for $U(r)$ that is applicable to most of
the nuclei. The primary choice is the conventional Wood-Saxon form commonly
used in the local optical model calculations:
\begin{equation}
U(x) = -\,\left(V_r\,f_r(x)\,+\,i W_i\,f_i(x)\,+\,i W_d\,f_d(x)
\right)
\label{eq8}
\end{equation}
\begin{equation}
\eqalign{
{\rm with}\,\,\,f_y(x) = \left[1\,+\,{\rm exp}\left(\frac{x-R_y}{a_y}
\right)\right]^{-1}\,\,\,\,y\,=\,r\,\,{\rm and}\,\,i
\cr
~~~~~~~f_d(x) = 4\,{\rm exp}\left(\frac{x-R_d}{a_d}\right)\,
\left[1\,+\,{\rm exp}\left(\frac{x-R_d}{a_d}\right)\right]^{-2}
}
\label{eq9}
\end{equation}

Recently Tian, Pang and Ma \cite{tpm15} have obtained a new set of
parameters for this type of potential by fitting the nucleon scattering
data on nuclei ranging from $^{27}$Al to $^{208}$Pb with incident
energies around 10 MeV to 30 MeV. It provides an excellent agreement
with large amount of cross section data and is energy independent. The
numerical values of these parameters can be found in Table~2 of
Ref.\cite{tpm15}. Henceforth, the potential obtained by this
parameterization will be referred to as ``TPM15". 

\subsection{The nonlocal kernel}
\begin{figure*}[htb!]
\centering
\subfigure[][Real part]{
\centering
\includegraphics[scale=0.43]{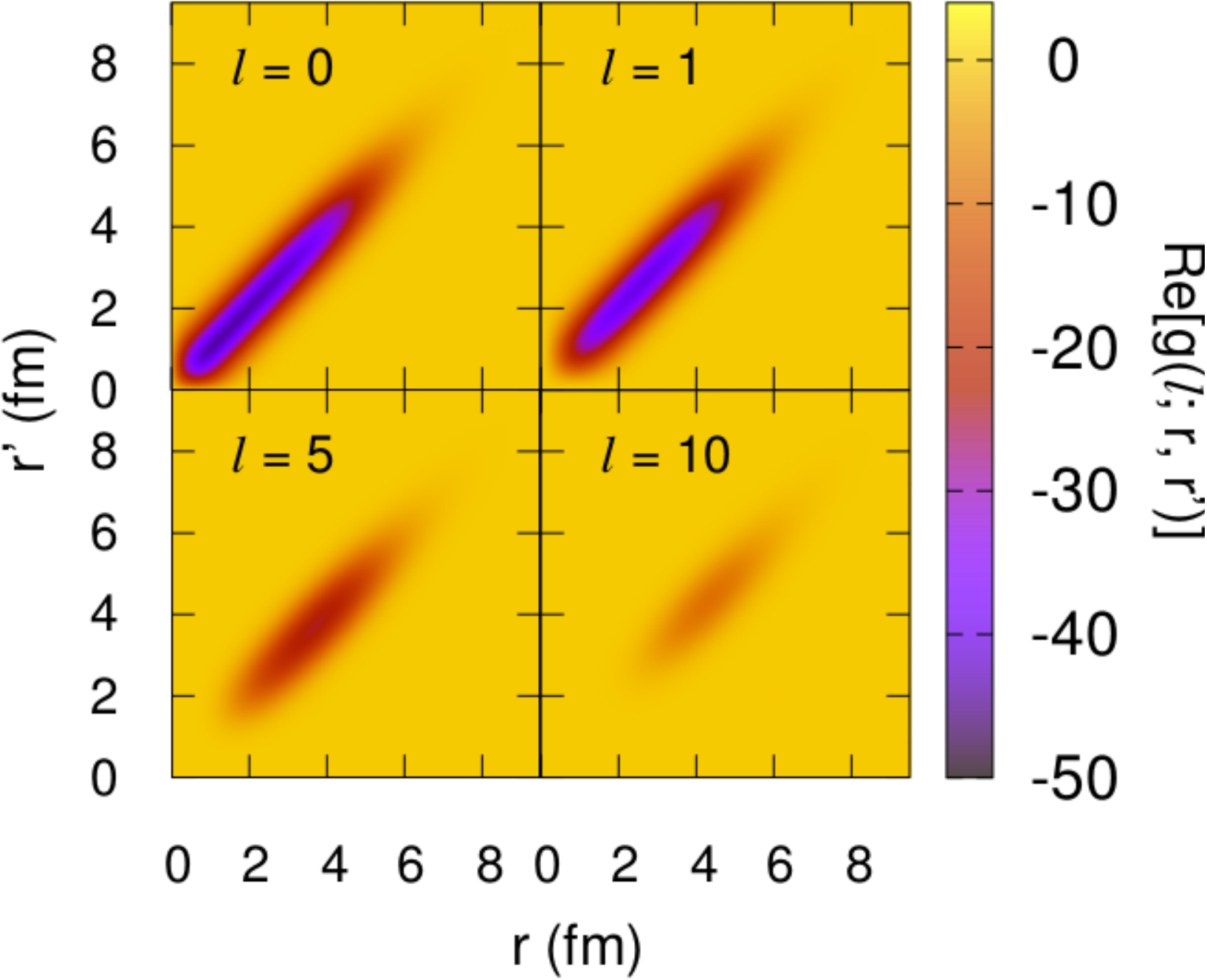}}
\subfigure[][Imaginary Part]{
\centering
\includegraphics[scale=0.43]{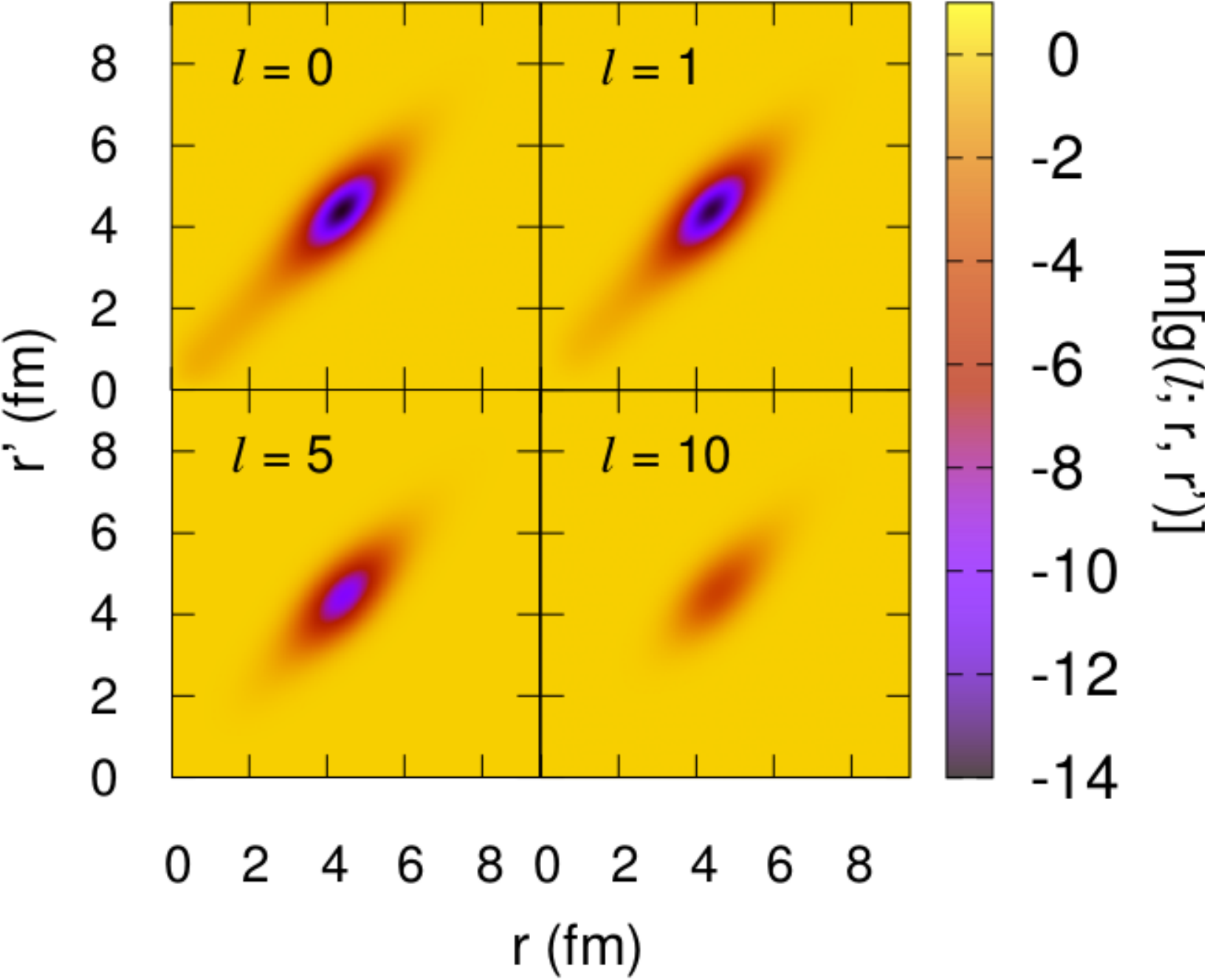}}
\caption{Behaviour of the nonlocal kernel as a function of relative
coordinates $r$ and $r^\prime$ for $l$ = 0, 1, 5 and 10. Calculations
are done with TPM15 parameterization for neutron scattering off $^{56}$Fe 
nucleus \cite{tpm15}. The non-local range used in calculations is 
$\beta$=0.90 fm.}
\label{f1}
\end{figure*}

We now examine the behaviour of the nonlocal kernel, $g(l;r,r^\prime)$,
appearing in the nonhomogeneous term in Eq.(\ref{eq5}). To illustrate,
we plot $g(l;r,r^\prime)$ as a function of $r$ and $r^\prime$ for different 
partial waves in Fig.\ref{f1} for neutron-$^{56}$Fe scattering. As it can
be seen, $g(l;r,r^\prime)$ is a well-behaved function which is symmetric
around $r$=$r^\prime$. Its strength diminishes with increasing $l$ and so
does the importance of the nonlocality.
This behaviour of $g(l;r,r^\prime)$ prompts us to use the mean
value theorem (MVT) of integral calculus \cite{mvt} to solve the 
integro-differential equation (Eq.(\ref{eq5})).

\subsection{The MVT technique}
According to the mean value theorem of integral calculus \cite{mvt}, if
a function $q(x)$ is non-negative and integrable on $[a,b]$ and $p(x)$
is continuous on $[a,b]$, then there exists $c \in [a,b]$ such that
\begin{equation}
\int_a^b p(x) q(x) dx = p(c) \int_a^b q(x) dx\,.
\label{eq10}
\end{equation}
The theorem holds for non-positive $q(x)$ as well.

\begin{figure*}[htb!]
\centering
\includegraphics[scale=0.43]{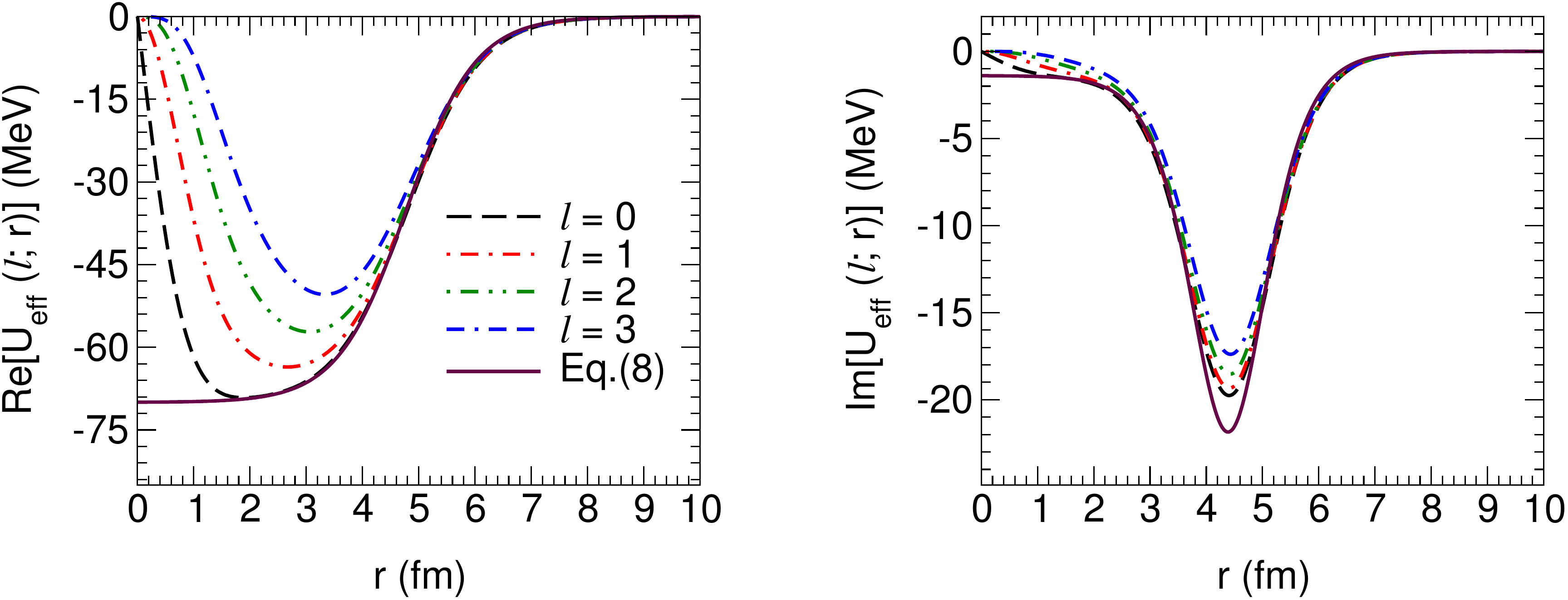}
\caption{Behaviour of $U_{\rm eff}(l;r)$ as a function of distance 
for $l$ = 0, 1, 2 and 3. Calculations are done with TPM15 parameterization
for n-$^{56}$Fe scattering \cite{tpm15}. The non-local range used in
calculations is $\beta$=0.90 fm.}
\label{f2}
\end{figure*}

We examine the applicability of this theorem to the kernel in Eq.(\ref{eq5}).
The integrand in Eq.(\ref{eq5}) is a product of $g(l;r,r^\prime)$ and
the wave function, $u(l;r^\prime)$. The wave function is continuous in
the interval $[0, r_m]$. The analytic structure of $g(l;r,r^\prime)$ 
makes it evident that the kernel is integrable. Considering the behavior
of $g(l;r,r^\prime)$ from Fig.\ref{f1}, the integral in Eq.(\ref{eq5})
can be written as
\begin{equation}
\int_0^{r_m}g(l;r,r^\prime)u(l;r^\prime)\,dr^\prime =
u(l;\xi)\int_{0}^{r_m} g(l;r,r^\prime)\,dr^\prime
\label{eq11}
\end{equation}
with $\xi \in [0, r_m]$. Further, from Fig.\ref{f1} it can be seen that 
$g(l;r,r^\prime)$ is strongly peaked at $r$=$r^\prime$ and is symmetric 
around it. With this observation, we can expand $u(l;\xi)$ about 
$r$=$r^\prime$. The leading term $u(l;r)$, evidently, is the most dominant
in the expansion. Therefore, we choose $u(l;\xi) = u(l;r)$, yielding
\begin{equation}
\int_0^{r_m} g(l;r,r^\prime) u(l;r^\prime) dr^\prime \approx u(l;r)
\int_0^{r_m} g(l;r,r^\prime) dr^\prime .
\label{eq12}
\end{equation}
This leads to the homogenized form of the Schr\"{o}dinger equation
\begin{equation}
\left[\frac{d^2}{dr^2}-\frac{l(l+1)}{r^2} + 
\frac{2\mu E}{\hbar^2}\right]u(l;r)\,=\,\frac{2\mu U_{\rm eff}(l;r)}
{\hbar^2}u(l;r)\,,
\label{eq13}
\end{equation}
where the effective local potential, $U_{\rm eff}(l;r)$, is given by
\begin{equation}
U_{\rm eff}(l;r) = \int_0^{r_m}\,g(l;r,r^\prime)\,dr^\prime .
\label{eq14}
\end{equation}
This potential contains the most dominant effect of nonlocality. It is
independent of energy, but depends on $l$. In Fig.\ref{f2} we show
$U_{\rm eff}(l;r)$ in comparison with the TPM15 potential (Eq.(\ref{eq8}))
for neutron-$^{56}$Fe system. It is observed that $U_{\rm eff}(l;r)$
gets reduced in strength as well as modified in shape. This is unlike
the local equivalent potential in Perey and Buck's work \cite{pb}.

\section{Accuracy of the method}
The homogenized Schr\"{o}dinger equation (Eq.(\ref{eq13})) obtained for
the description of Eq.(\ref{eq5}), of course, is very neat and useful,
but it has an approximation of calculating the kernel at $r$=$r^\prime$.
This needs to be tested carefully. For this purpose we need to compare
the solution of the homogeneous equation with that of the original 
integro-differential equation (Eq.(\ref{eq5})). This is achieved by solving
Eq.(\ref{eq5}) using an iterative scheme. This scheme is initiated by
the solution of Eq.(\ref{eq13}) using a suitable boundary condition.
The subsequent higher order solutions are obtained with the help of the
following iterative scheme:
\begin{eqnarray}
\hspace{-2cm}&&
\left[\frac{d^2}{dr^2}-\frac{l(l+1)}{r^2} + \frac{2\mu E}{\hbar^2}
- \frac{2\mu\,U_{\rm eff}(l;r)}{\hbar^2}\right]u_{i+1}(l;r)
\label{eq15}
\\
\nonumber
&&\hspace{0.5cm}=
\frac{2\mu}{\hbar^2}\int_0^{r_m} g(l;r,r^\prime)
u_i(l;r^\prime)\,dr^\prime - \frac{2\mu U_{\rm eff}(l;r)}{\hbar^2}
u_i(l;r),\,\,\,\,\,({\rm for\,\,all}\,\,i \ge 0)
\end{eqnarray}
The iterations are continued till the absolute value of difference
between the logarithmic derivatives of the wave function at the matching
radius in the $i^{\rm th}$ and the $(i+1)^{\rm th}$ steps is less
than or equal to 10$^{-6}$.

\begin{figure*}[htb!]
\centering
\subfigure[][n-$^{12}$C scattering]{
\centering
\includegraphics[scale=0.43]{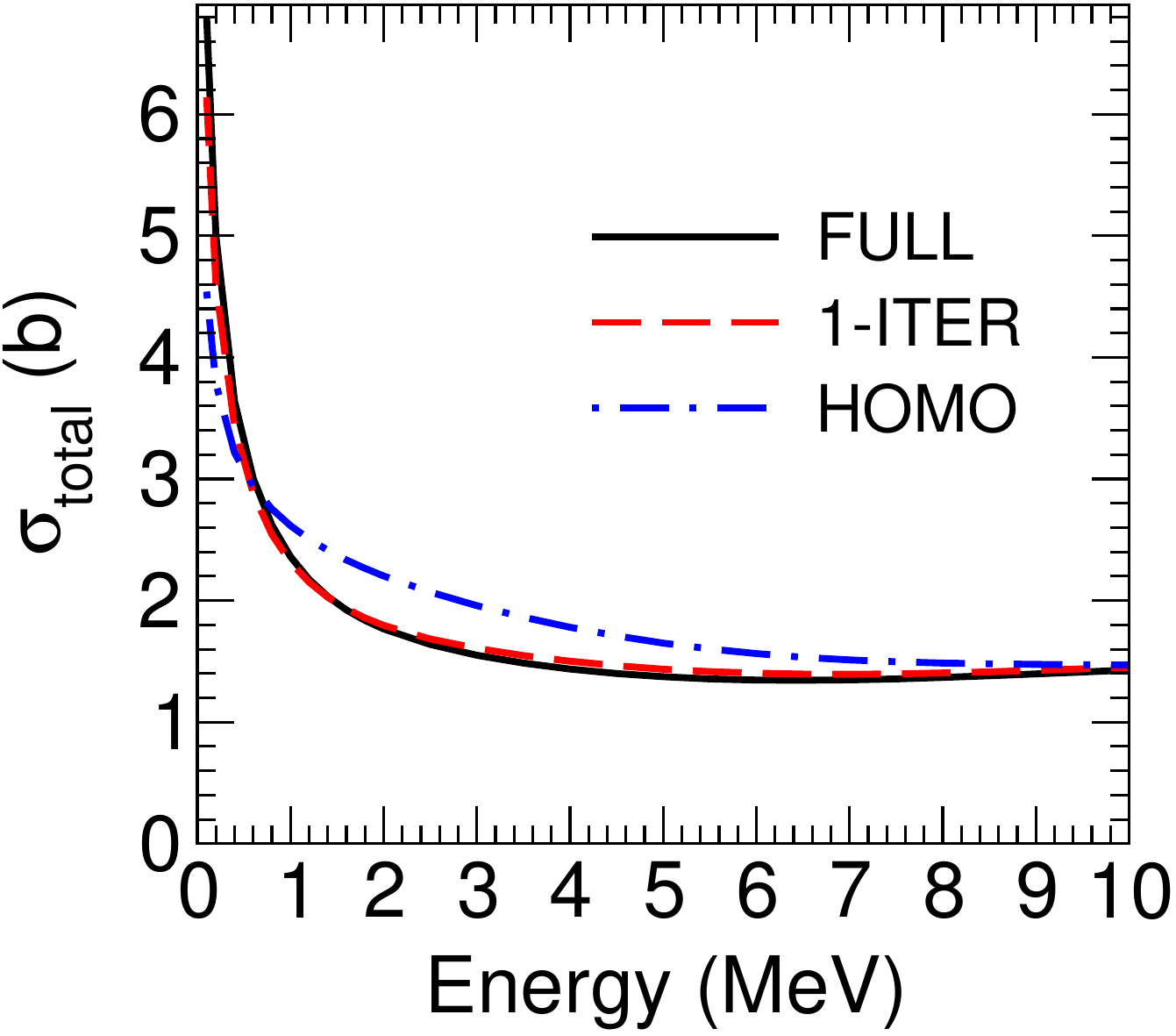}}
\hspace{1.5cm}
\subfigure[][n-$^{56}$Fe scattering]{
\centering
\includegraphics[scale=0.43]{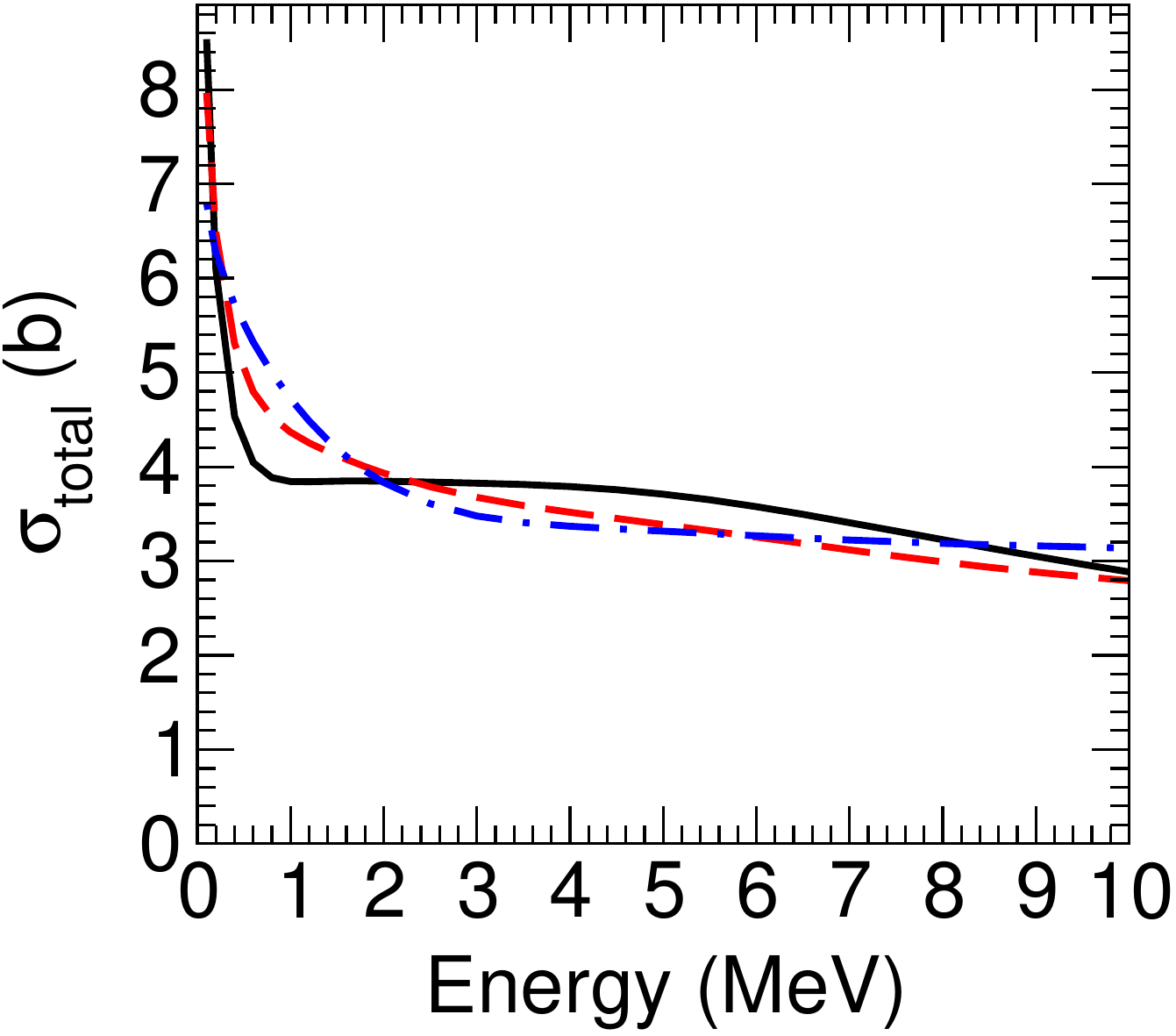}}
\caption{Total cross sections for neutron scattering off $^{12}$C 
and $^{56}$Fe nuclei. Calculations are done with TPM15 parameterization
\cite{tpm15}.}
\label{f3}
\end{figure*}

\begin{figure*}[htb!]
\centering
\subfigure[][n-$^{12}$C scattering]{
\centering
\includegraphics[scale=0.4]{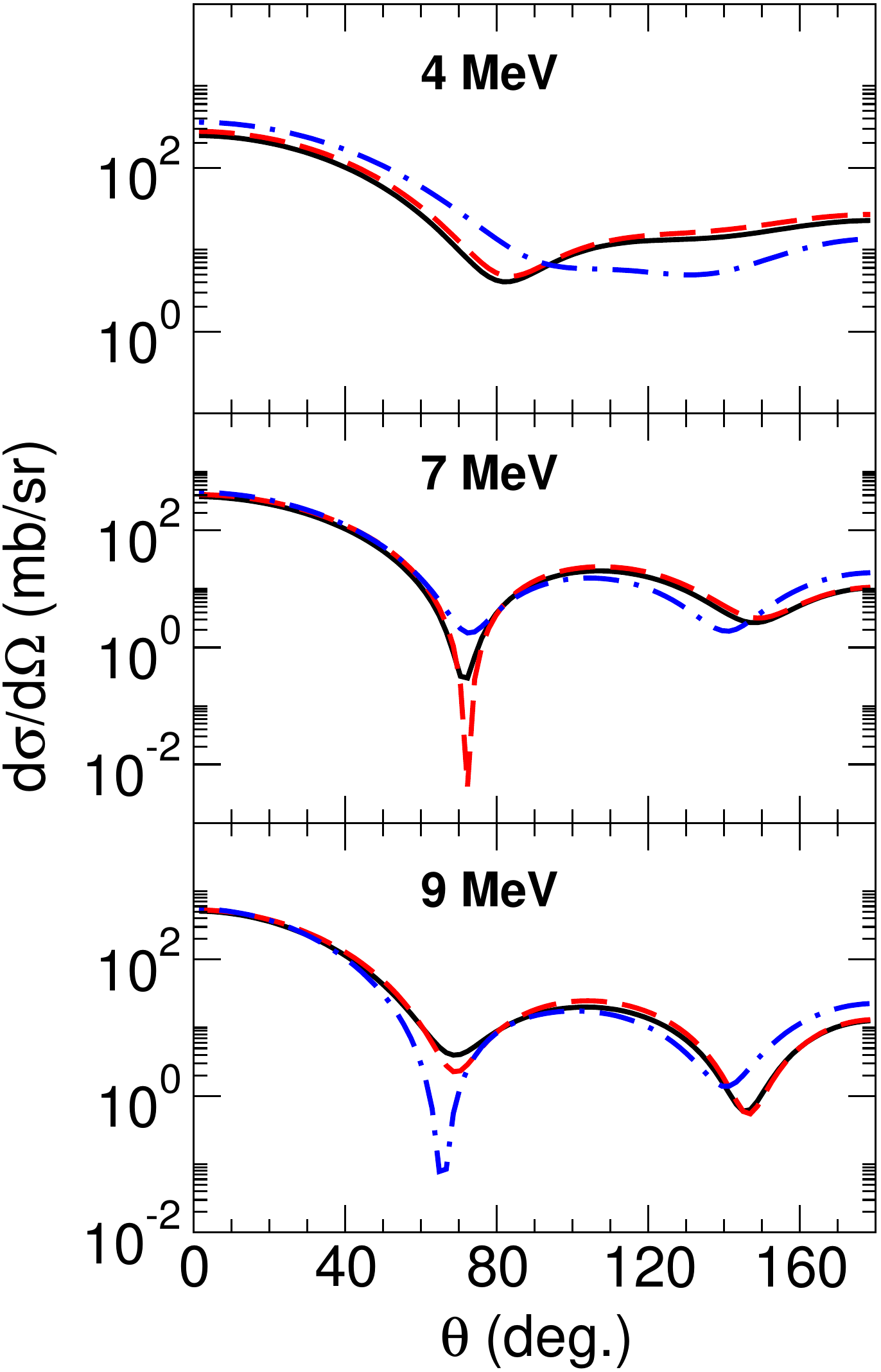}}
\hspace{1.5cm}
\subfigure[][n-$^{56}$Fe scattering]{
\centering
\includegraphics[scale=0.4]{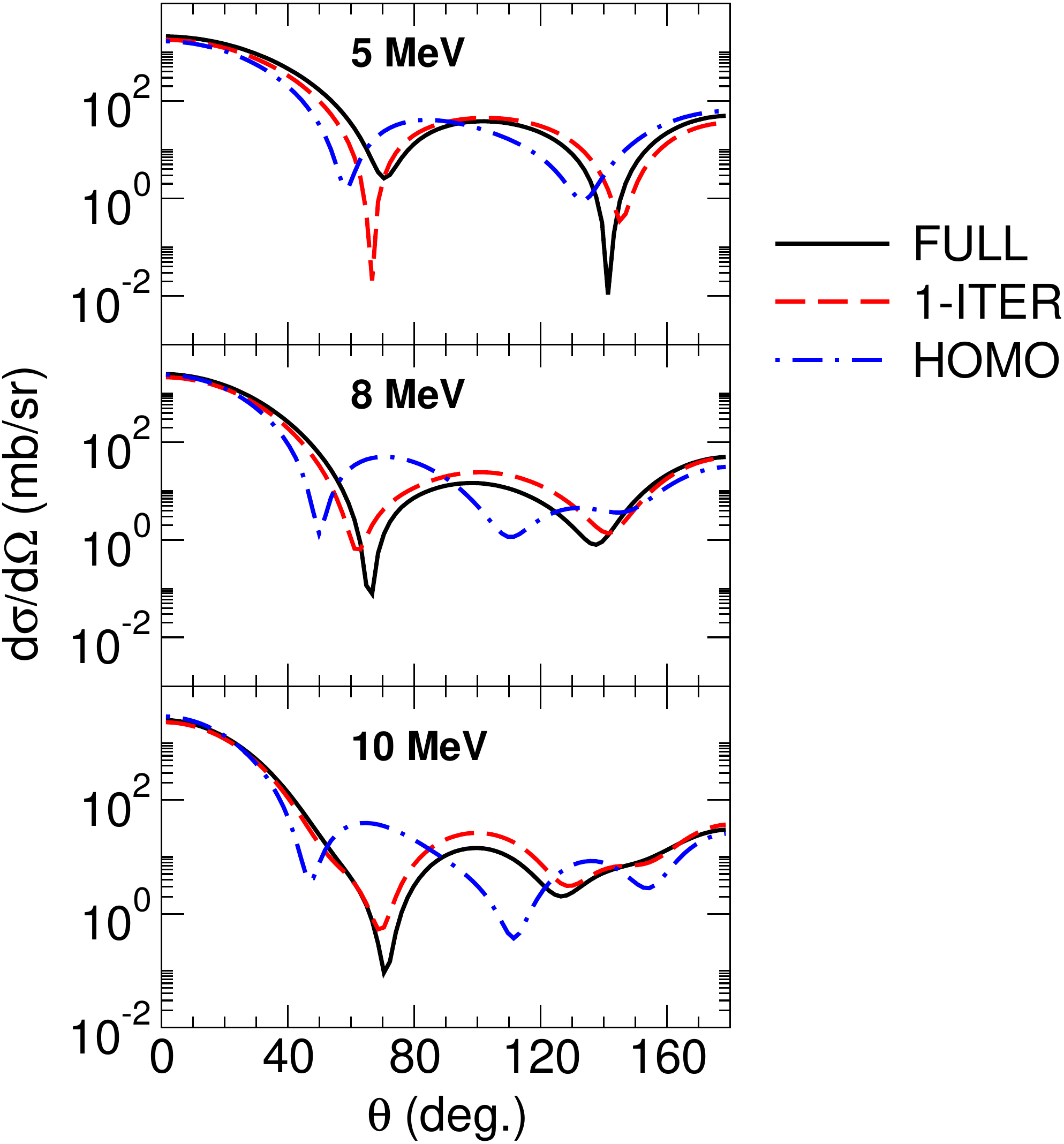}}
\caption{Angular distributions for neutron scattering off $^{12}$C
and $^{56}$Fe nuclei. Calculations are done with TPM15 parameterization
\cite{tpm15}.}
\label{f4}
\end{figure*}
In Figs.\ref{f3}-\ref{f4} we show the calculated total and differential
cross sections for n-$^{12}$C and n-$^{56}$Fe scatterings in the low energy
domain for various approximations to Eq.(\ref{eq5}). Results obtained by
solving Eq.(\ref{eq13}) and Eq.(\ref{eq15}) are labeled as HOMO and FULL 
respectively. To understand the impact of the iterative scheme, in 
Figs.\ref{f3}-\ref{f4} we also show results obtained after one iteration
(labeled as 1-ITER). We notice that though the homogeneous results have 
significant difference (especially for $^{56}$Fe) from the FULL results,
they are, overall, not far away from them. The 1-ITER results, however, 
are pretty close to the full results of Eq.(\ref{eq5}). This is very nice,
because in actual calculations one can then use just one iteration and 
get results very close to the FULL results. Computationally, once we 
have homogeneous results it is straight forward to get 1-ITER results.

\subsection{Dependence on the choice of $U(r)$}
To examine further the dependence of the accuracy of our method on the
choice of $U(r)$, we perform above calculations for another potential.
We construct $\displaystyle{U(r)=V(r)+i W(r)}$ such that
$V(r)$ is obtained microscopically through folding model \cite{satlov}: 
\begin{equation}
V(r)\,=\,\int d{\bf r}_2\,\rho({\bf r}_2)\,v(r_{12})\,,
\label{eq16}
\end{equation}
where $v(r_{12})$ is the effective nucleon-nucleon interaction,
$\rho({\bf r}_2)$ is the total nucleon density of the target and $r_{12}$
is the distance between the projectile (neutron) and a nucleon in the
target nucleus. To make it a reasonable representation for the
present study, TPM15 parameterization (see Eqs.(\ref{eq8})-(\ref{eq9}))
is used for $W(r)$. This prescription of $U(r)$ will be referred to as 
``FoldTPM15" in subsequent discussions.

For the effective nucleon-nucleon interaction we have
used the well known M3Y prescription \cite{m3y}:
\begin{equation}
v(r)\,=\,\left[7999\,\frac{e^{-4r}}{4r}\,-\,2134\,\frac{e^{-2.5r}}{2.5r}
\right]\,\,{\rm MeV},
\label{eq17}
\end{equation}
where the two Yukawa terms represent the direct contribution of the 
interaction. In principal, one needs to take into account the knock-on
contribution as well \cite{satlov}. Since the knock-on contribution in the
nucleon-nucleus scattering is expected to be insignificant in the 
low energy range \cite{lemere}, we have not included it in Eq.(\ref{eq17}).
The density distribution of target nucleus has been calculated using
the well established relativistic mean field model \cite{YKG.90},
which is known to reproduce ground state properties of nuclei spanning
the entire periodic table.
\begin{figure*}[th!]
\centering
\subfigure[][The nonlocal kernel]{
\centering
\includegraphics[scale=0.43]{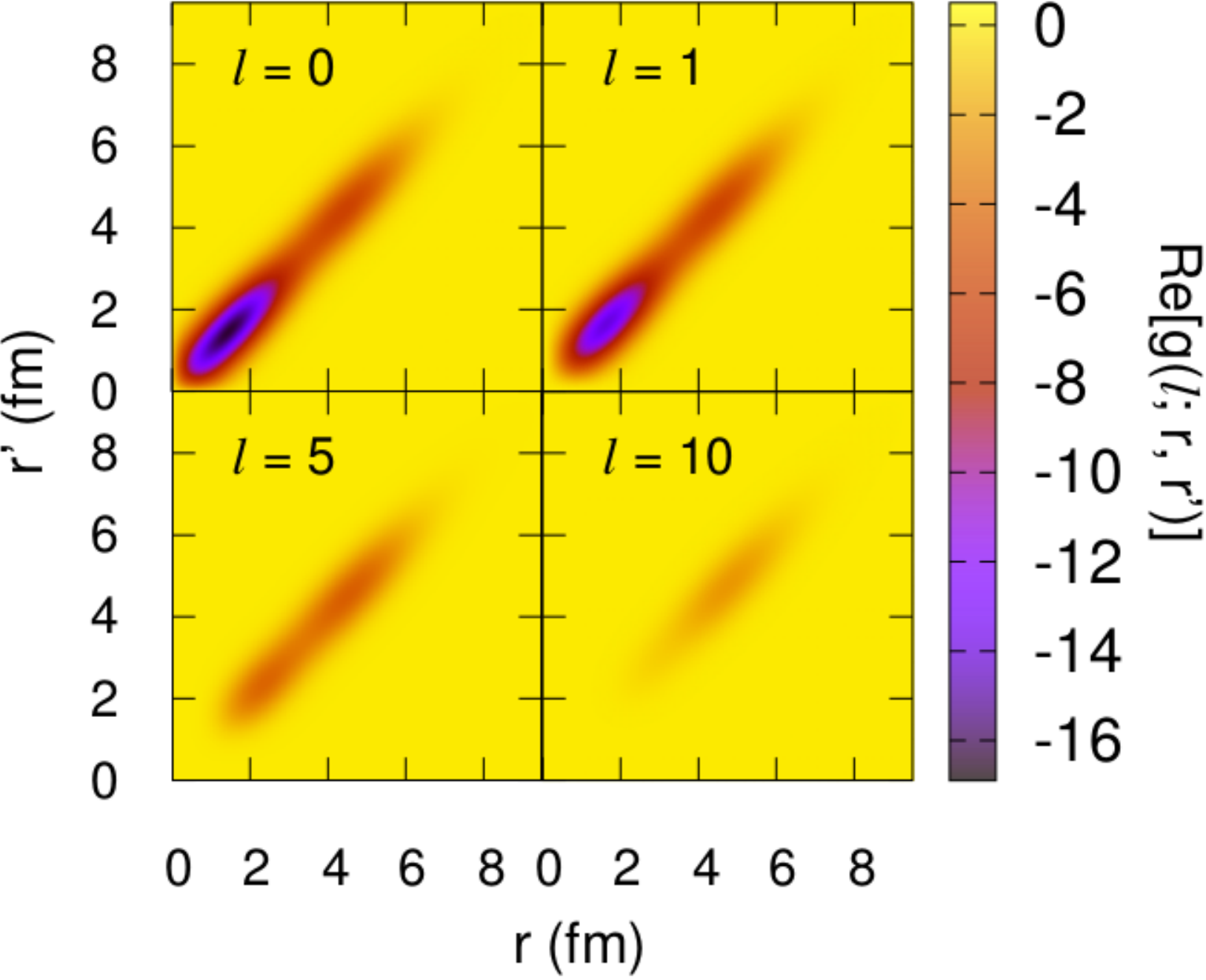}
\label{f5a}}
\hspace{1.0cm}
\subfigure[][$U_{\rm eff}(l;r)$]{
\centering
\includegraphics[scale=0.41]{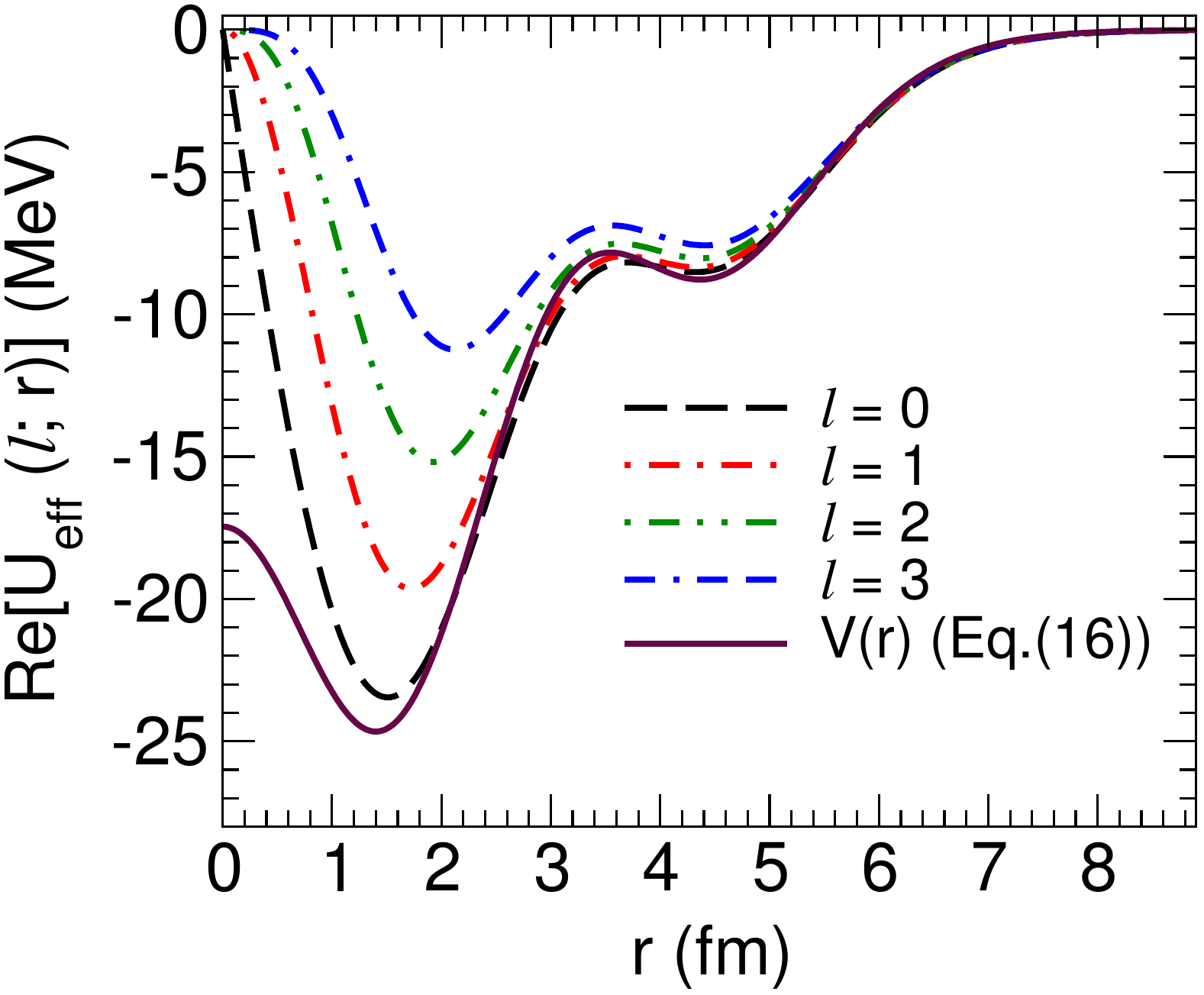}
\label{f5b}}
\caption{Behaviour of the real part of the nonlocal kernel and
$U_{\rm eff}(l;r)$ for neutron scattering off $^{56}$Fe nucleus.
Calculations are done with FoldTPM15 prescription. The non-local
range used in calculations is $\beta$=0.90 fm.}
\label{f5}
\end{figure*}

In Fig.\ref{f5a}, we show the behaviour of the real part of the nonlocal
kernel function calculated using FoldTPM15 prescription for neutron 
scattering off $^{56}$Fe. It has the required structure of being well
behaved and symmetric about $r$=$r^\prime$ for the applicability
of the MVT technique. We also show the corresponding real part of 
$U_{\rm eff}(l;r)$ in Fig.\ref{f5b}.

Comparing the contour plot for $g(l;r,r^\prime)$ (see Fig.\ref{f5a})
with the earlier plot in Fig.\ref{f1} for TPM15, we also notice that the
effect of nonlocality in it probably is much less.
 
\begin{figure*}[htb!]
\centering
\subfigure[][n-$^{12}$C scattering]{
\centering
\includegraphics[scale=0.43]{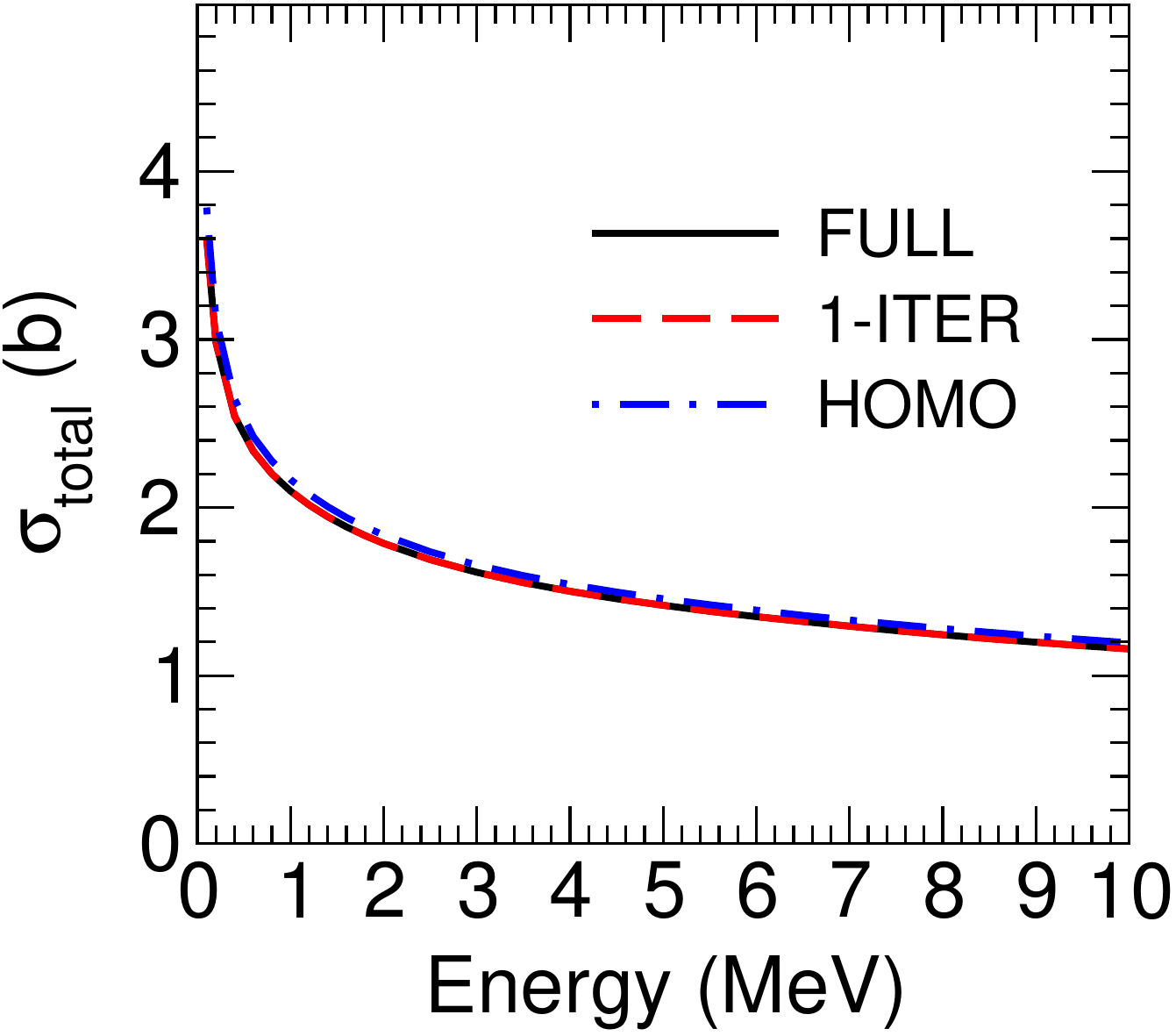}}
\hspace{1.5cm}
\subfigure[][n-$^{56}$Fe scattering]{
\centering
\includegraphics[scale=0.43]{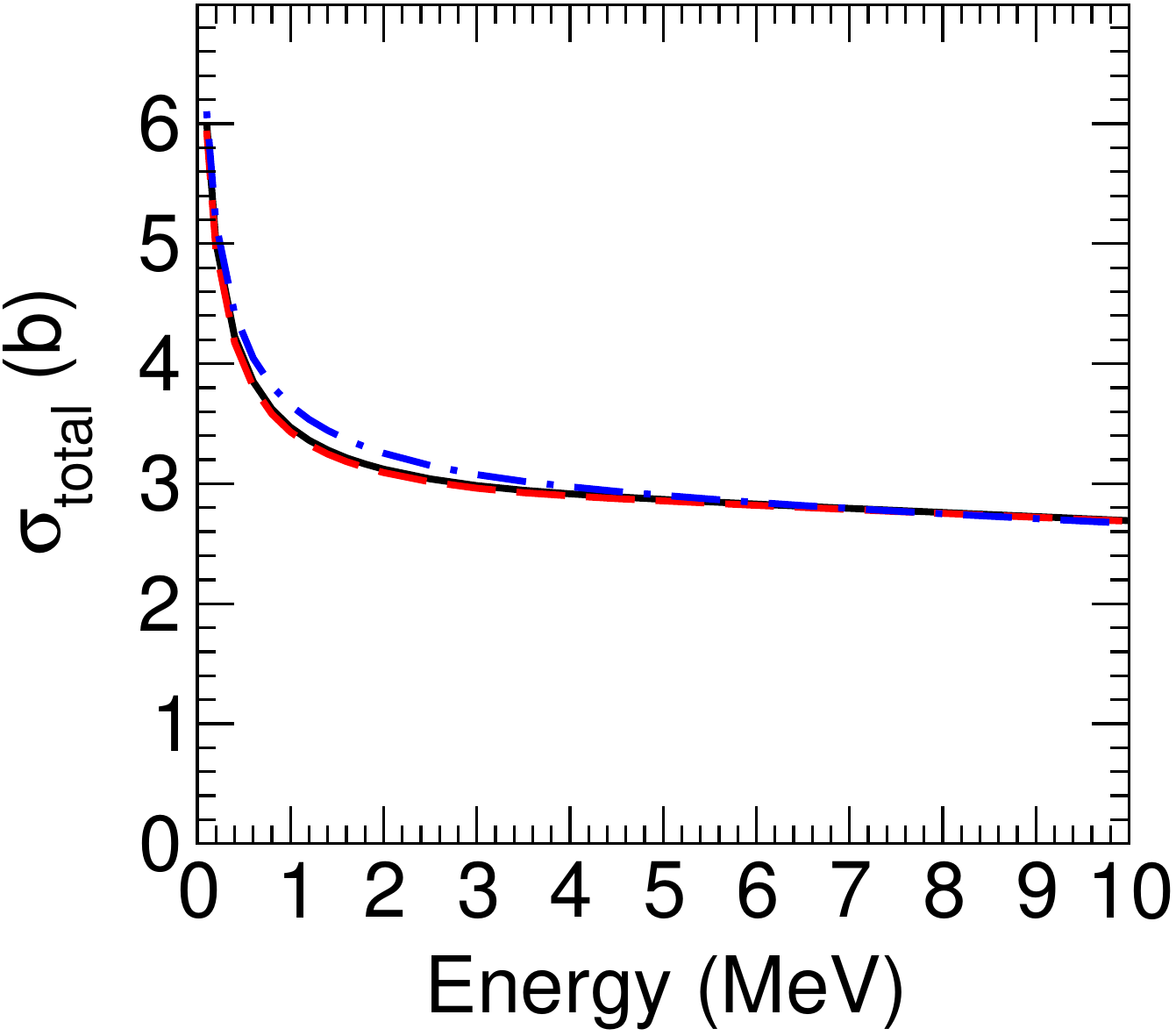}}
\caption{Total cross sections for neutron scattering off $^{12}$C and
$^{56}$Fe nuclei. Calculations are done with FoldTPM15
prescription.}
\label{f6}
\end{figure*}

\begin{figure*}[htb!]
\centering
\subfigure[][n-$^{12}$C scattering]{
\centering
\includegraphics[scale=0.4]{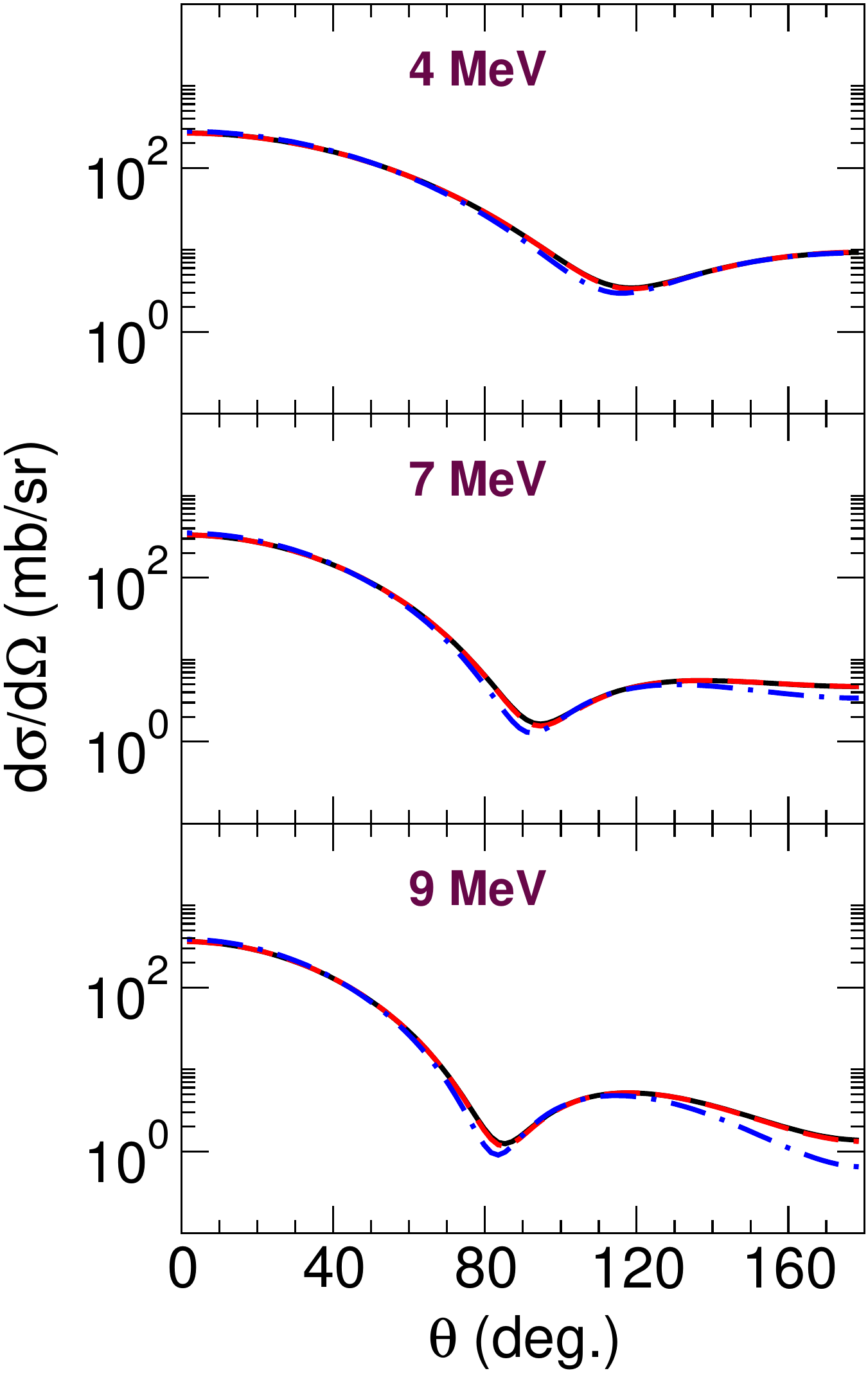}}
\hspace{1.5cm}
\subfigure[][n-$^{56}$Fe scattering]{
\centering
\includegraphics[scale=0.4]{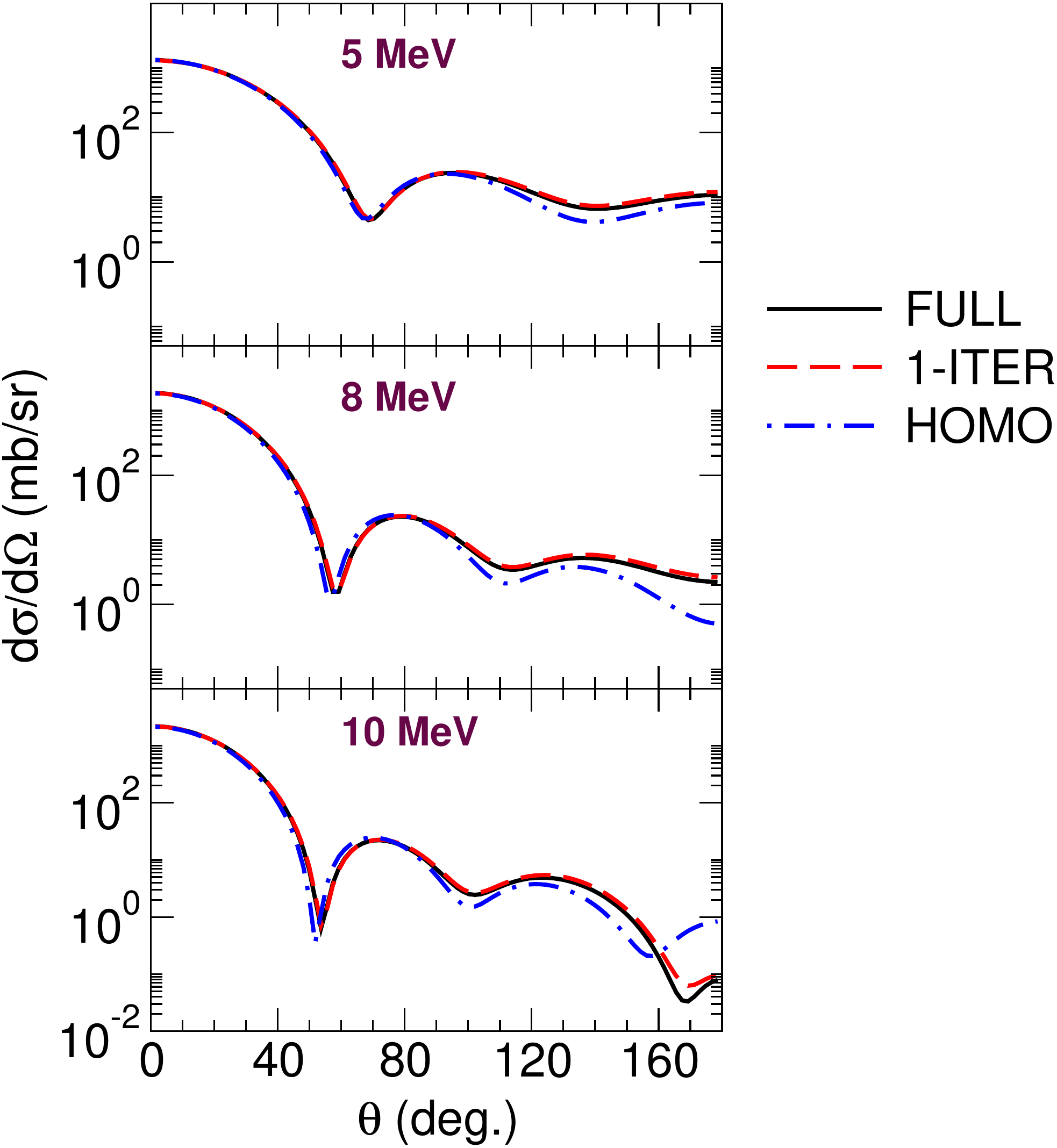}}
\caption{Angular distributions for neutron scattering off $^{12}$C and
$^{56}$Fe nuclei. Calculations are done with FoldTPM15
prescription.}
\label{f7}
\end{figure*}

The calculated total and differential cross sections shown in 
Figs.\ref{f6}-\ref{f7} are for (i) solution of the homogeneous equation, 
(ii) solution obtained after one iteration and (iii) converged solution of 
Eq.(\ref{eq15}) for n-$^{12}$C and n-$^{56}$Fe scatterings. The conclusion
about the accuracy of our technique in this case, if anything, is better 
than that seen in Figs.\ref{f3}-\ref{f4}. Thus, we conclude that the 
accuracy of our technique is good. This improvement in accuracy, as
mentioned above, may be due to lesser effect of nonlocality in FoldTPM15.

\subsection{Impact of approximation leading to Eq.({\ref{eq7}})}
\begin{figure*}[htb!]
\centering
\subfigure[][n-$^{12}$C scattering]{
\centering
\includegraphics[scale=0.43]{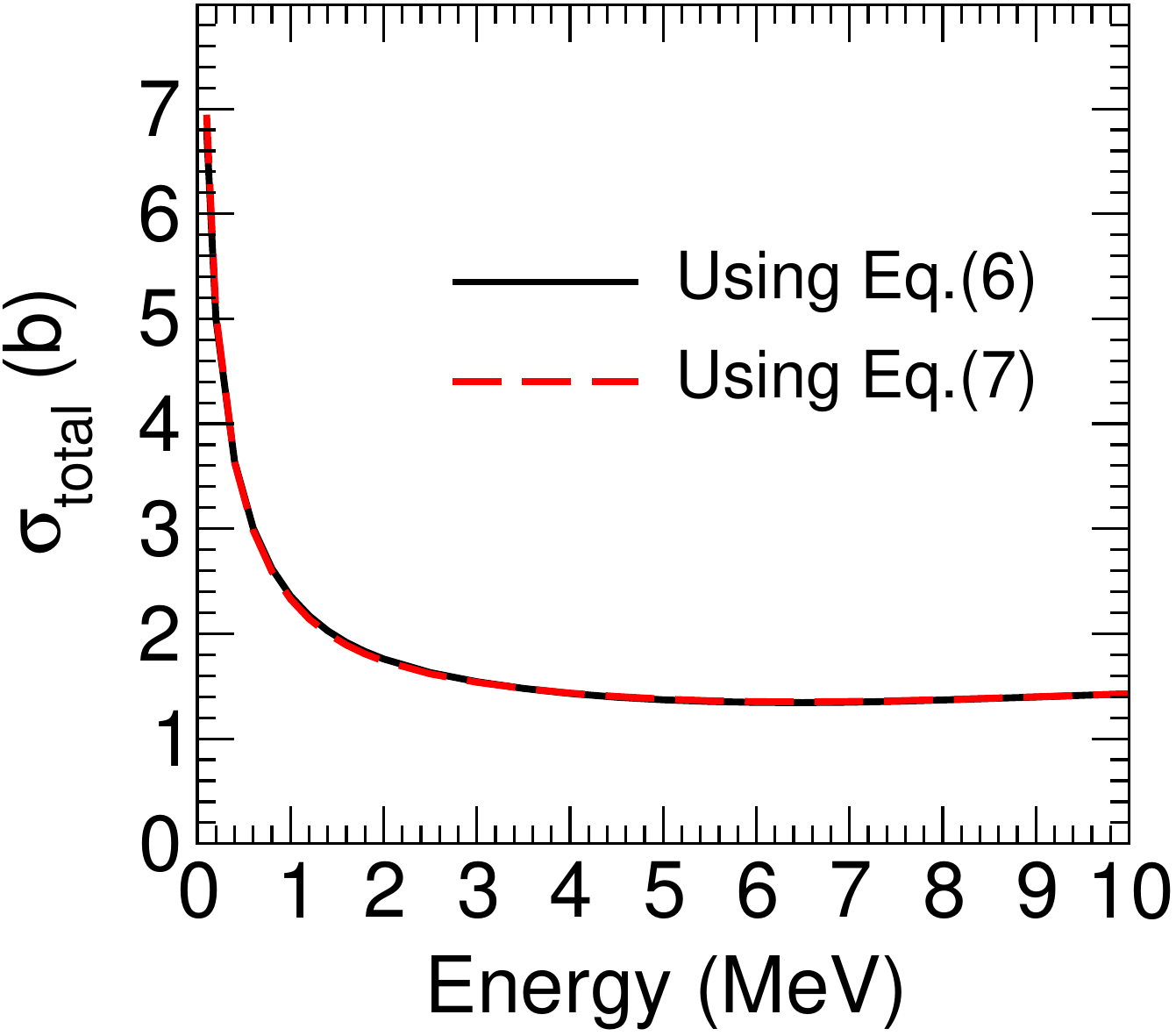}}
\hspace{1.5cm}
\subfigure[][n-$^{56}$Fe scattering]{
\centering
\includegraphics[scale=0.43]{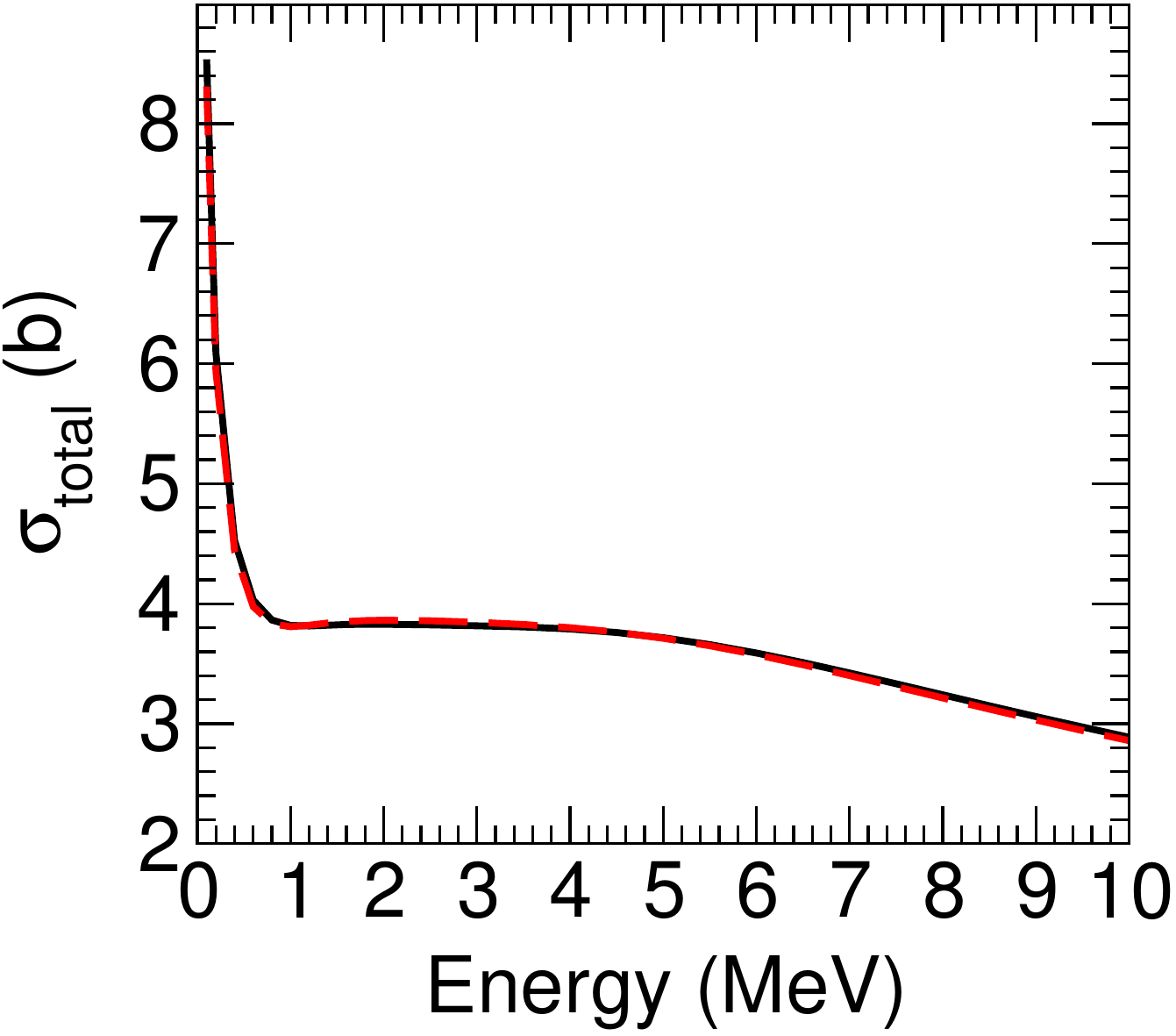}}
\caption{Total cross sections calculated using Eqs.(\ref{eq6})-(\ref{eq7})
for neutron scattering off $^{12}$C and $^{56}$Fe nuclei. Calculations
are done with TPM15 parameterization \cite{tpm15}.}
\label{f8}
\end{figure*}

As mentioned in Section 2, the nonlocal kernel represented by Eq.(\ref{eq6})
contains the nonlocal potential, $\displaystyle{U\left(\frac{|\bf{r}+
\bf{r^\prime}|}{2}\right)}$ inside the integrand. Common practice is
to approximate $|{\bf r}+{\bf r^\prime}|$ by $(r+r^\prime)$ leading to
Eq.(\ref{eq7}). We study the impact of this approximation on the accuracy
of results obtained by solving Eq.(\ref{eq15}). In Fig.\ref{f8}, we
compare the total cross sections obtained by using Eq.(\ref{eq6}) in 
Eq.(\ref{eq15}) with those obtained by using Eq.(\ref{eq7}). Results
are shown for neutron scattering off $^{12}$C and $^{56}$Fe targets
using TPM15 parameterization. We find that use of Eq.(\ref{eq7}) in 
Eq.(\ref{eq15}) is good within 1$\%$ in the considered energy range.

\section{Impact of different choice of nonlocal form factor}
\subsection{Impact of the form of nonlocality}
Recently Rotureau {\it et al.} have developed a method to construct
nonlocal optical potential from first principle \cite{rot17}. In such
microscopic approach, since the nonlocality is inherent in the formalism,
its form could be different from a Gaussian. Therefore, to determine the
impact of different choices of nonlocal form factor, we take, in
addition to the Gaussian form used so far, an exponential form:
$\displaystyle{{\rm exp}\left(-|\vec{r}-\vec{r^\prime}|/\alpha\right)}$, 
which has normalization similar to that in Eq.(\ref{eq2}). Further, we
fix the value of $\alpha$ such that both the form factors have same rms
radius. Thus, we have two form factors with same normalization and
rms radius but different shapes. To see the effect of such pair of form
factors we solve the full scattering equation (Eq.(\ref{eq15})) with
these two form factors for $l$=0 and 1. We find that (i) the two wave
functions are very close to each other in magnitude and shape inside the
nucleus, and (ii) on the surface both of them have very close logarithmic
derivatives, which, as we know, determine the phase shifts. This allows
us to conclude that different form factors having same normalization and
same rms radius should give similar results for the scattering as well as
the reaction observables on a nucleus.

\subsection{Impact of the range of nonlocality}
\begin{figure*}[htb!]
\centering
\includegraphics[scale=0.43]{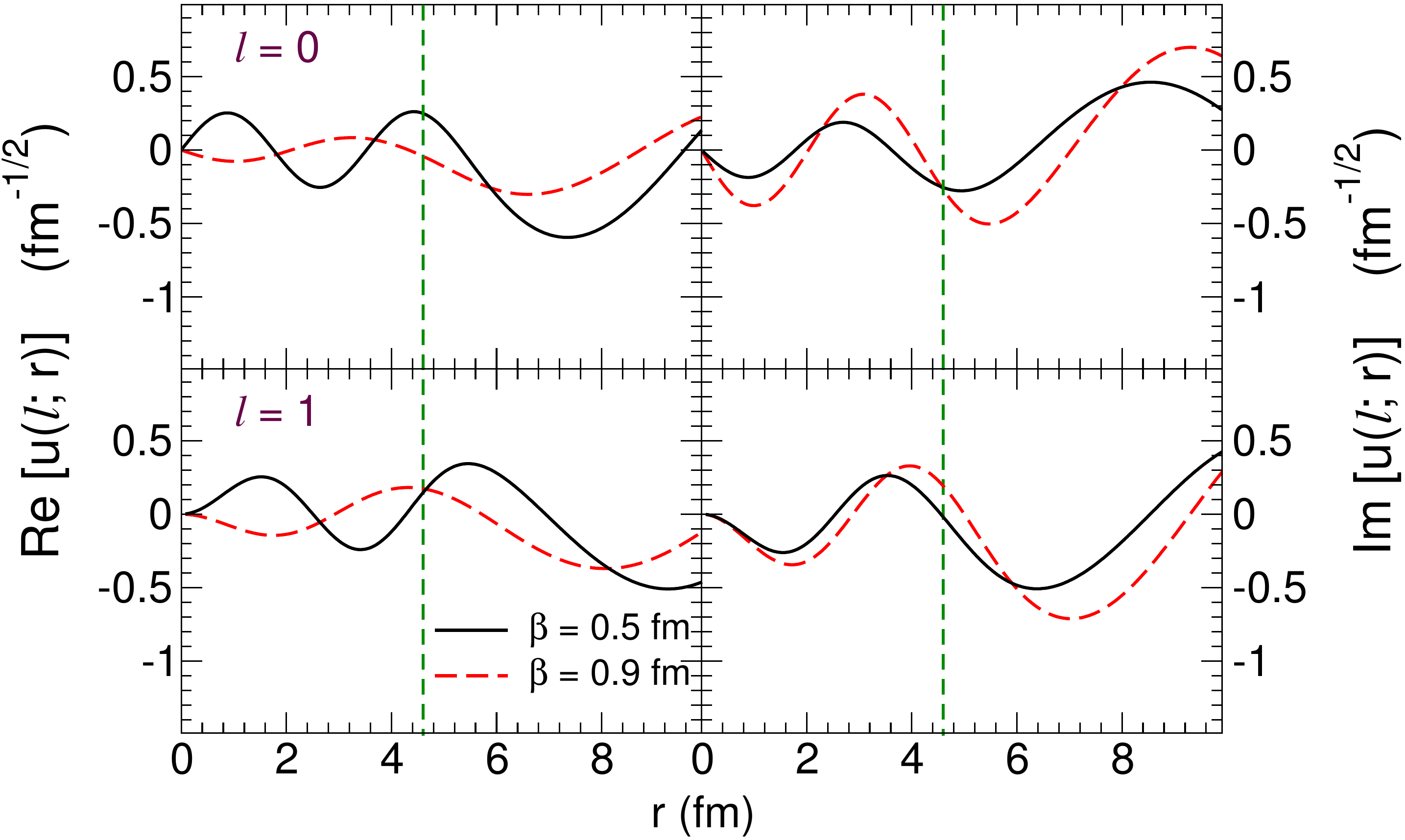}
\caption{Wave functions corresponding to two values of $\beta$ as a
function of distance. Results are shown for $l$=0 and 1 for
n-$^{56}$Fe scattering at 10 MeV. Calculations are done with TPM15
parameterization \cite{tpm15}. The vertical dashed line at 4.6 fm
marks the radius of $^{56}$Fe nucleus.}
\label{f9}
\end{figure*}

Next we explore the effect of different choices of rms radius for a
particular form factor. We take two values of $\beta$, {\it i.e.,} 0.9
and 0.5 fm, for the Gaussian form and plot in Fig.\ref{f9} wave functions
for $l$=0 and 1 for neutron-$^{56}$Fe scattering at 10 MeV. It is evident
that the wave functions change significantly with $\beta$ in the nuclear
interior and beyond. This amount of difference should be seen in nuclear
reactions as well.

\section{Comparison with experiments}

Having established the accuracy of our technique, we now present comparison
of the calculated observables with the data. However, now in the 
Schr\"{o}dinger equation we also include spin-orbit term. The resulting
Schr\"{o}dinger equation is:
\begin{equation}
\fl \left[\frac{d^2}{dr^2} - \frac{l(l+1)}{r^2} + 
\frac{2\mu\,E}{\hbar^2} + \frac{2\mu V_{SO}(r)}{\hbar^2} f_{jl}
\right] u(j;l;r) =\frac{2\mu}{\hbar^2}\int_0^{r_m} g(l;r,r^\prime) 
u(j;l;r^\prime) dr^\prime
\label{eq18}
\end{equation}
\begin{equation}
\hspace{-0.5cm}\eqalign{
{\rm where}\,\,\,\,f_{jl} = \frac{1}{2}\left(j(j+1)-l(l+1)-s(s+1)\right)
\,\, ({\rm with}\,\,s=1/2)
\cr
~~~~~~~~V_{SO}(r) = \left(U_{SO}\,+\,i W_{SO}\right) s(r)
\cr
~~~~~~~~s(r) = \left(\frac{\hbar^2}{{m_{\pi}}^2 c^2}\right)
\left(\frac{1}{a_{so} r}\right)
{\rm exp}\left(\frac{r-R_{so}}{a_{so}}\right)\left[1+
{\rm exp}\left(\frac{x-R_{so}}{a_{so}}\right)\right]^{-2}.
}
\label{eq19}
\end{equation}
For calculating $V_{SO}(r)$, TPM15 parameterization \cite{tpm15} is used.

\begin{figure*}[htb!]
\centering
\includegraphics[scale=0.40]{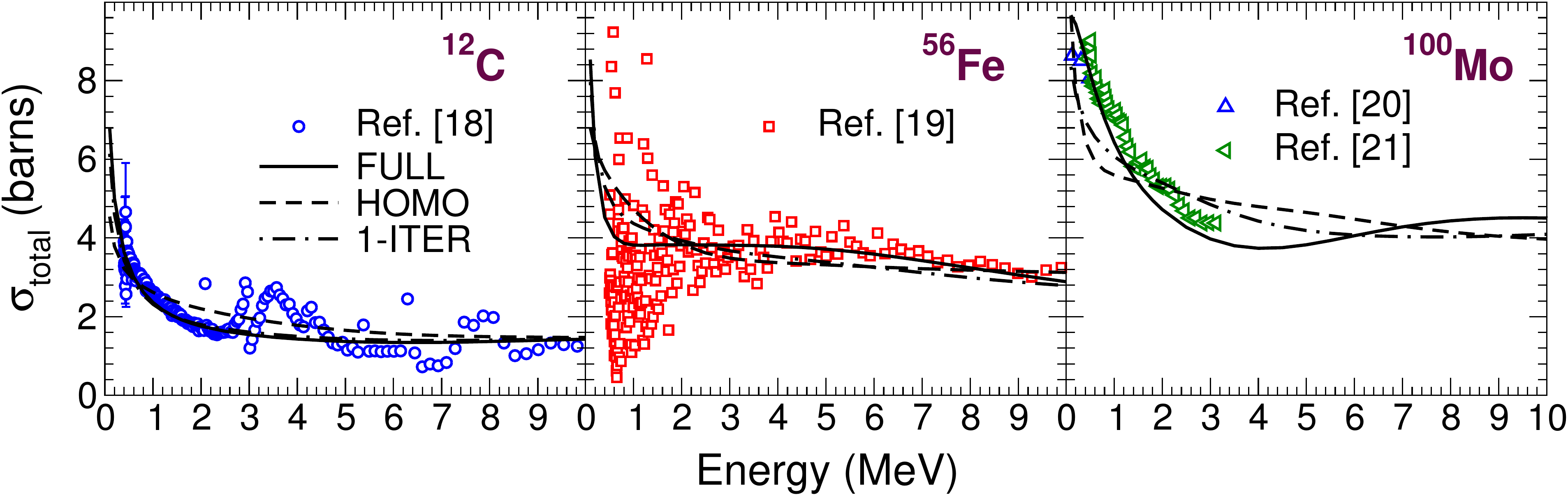}
\caption{Calculated total cross sections for neutron scattering off 
$^{12}$C, $^{56}$Fe and $^{100}$Mo nuclei along with the data 
\cite{rapp,geel,diva,pase}. Calculations are done with TPM15 
parameterization \cite{tpm15}.}
\label{f10}
\end{figure*}

\begin{figure*}[htb!]
\centering
\subfigure[][n-$^{12}$C scattering]{
\centering
\includegraphics[scale=0.36]{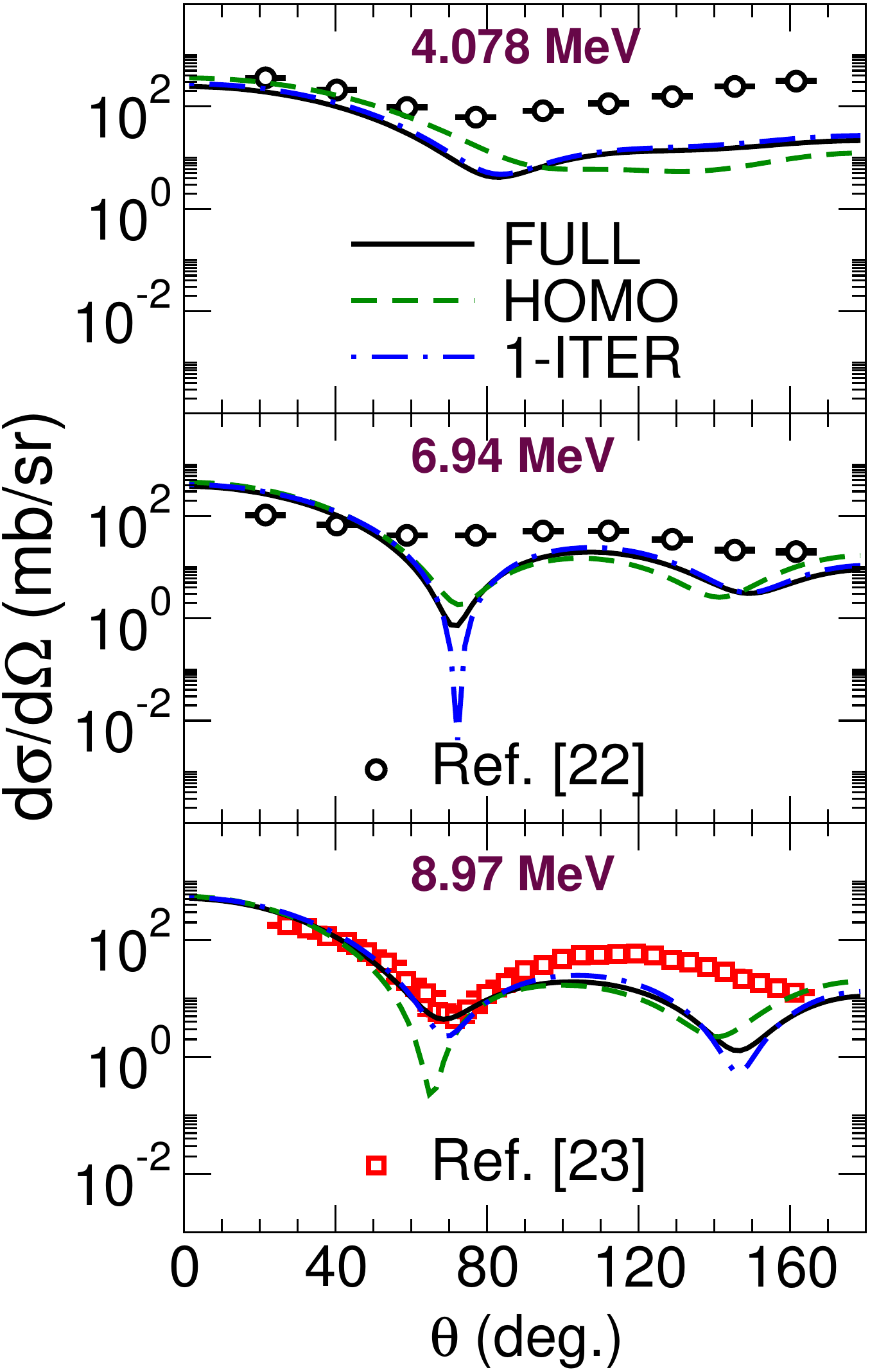}
\label{f11a}}
\subfigure[][n-$^{56}$Fe scattering]{
\centering
\includegraphics[scale=0.36]{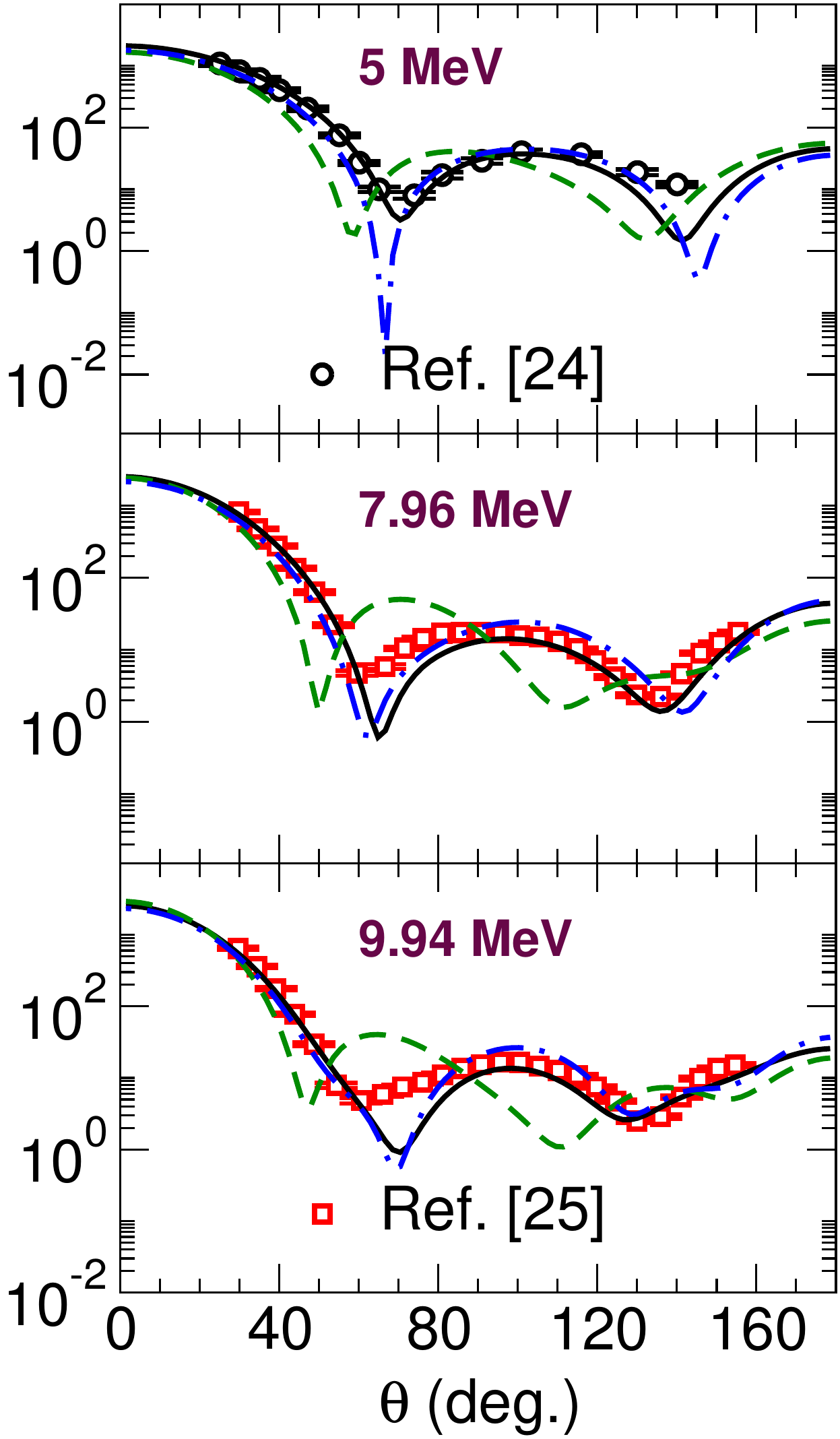}
\label{f11b}}
\subfigure[][n-$^{100}$Mo scattering]{
\centering
\includegraphics[scale=0.36]{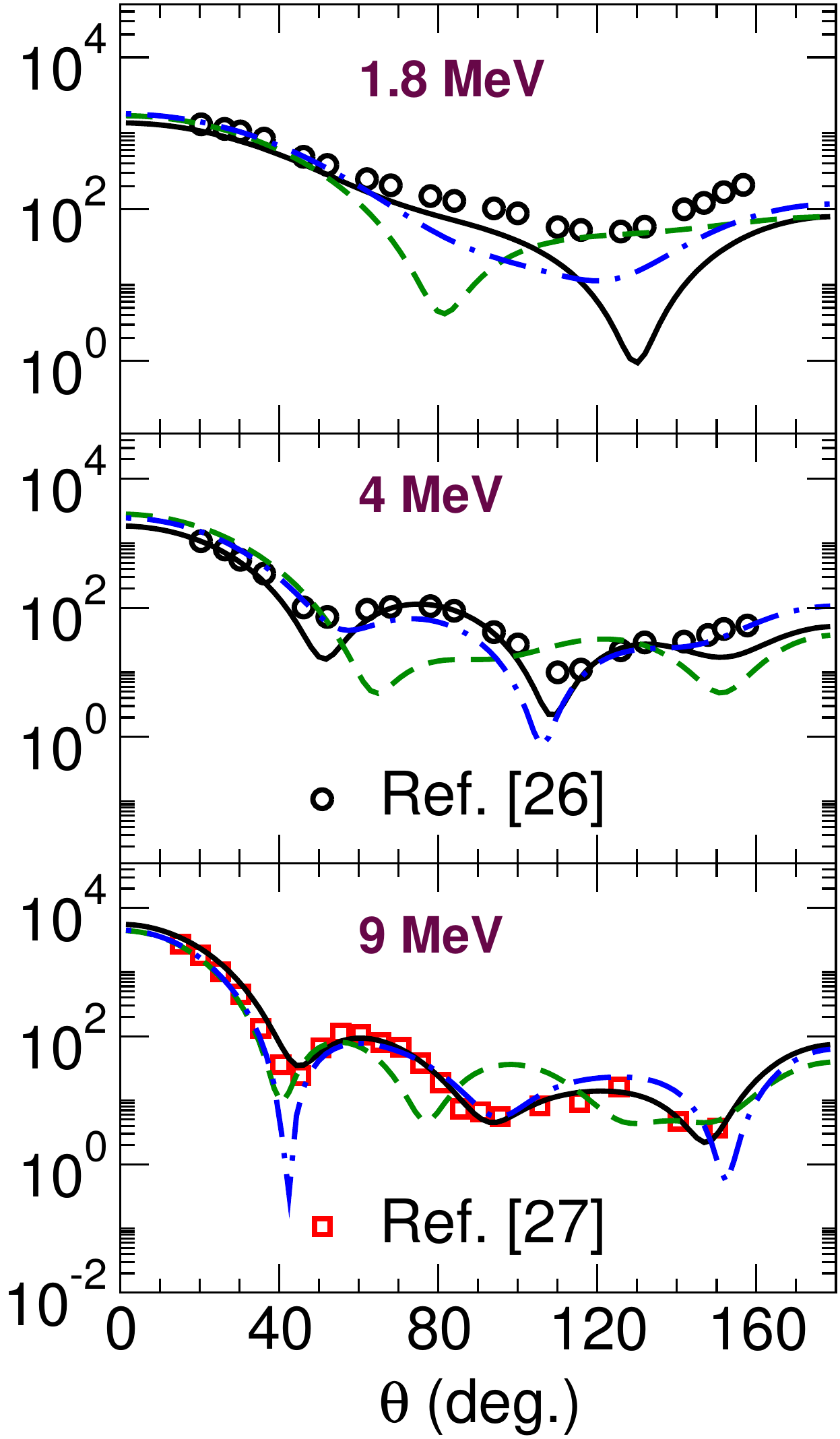}
\label{f11c}}
\caption{Calculated angular distributions using for neutron scattering
off $^{12}$C, $^{56}$Fe and $^{100}$Mo nuclei along with the data 
\cite{white,glas,kinney,kadi,smith,rapaport}. Calculations are done with
TPM15 parameterization \cite{tpm15}.}
\label{f11}
\end{figure*}

As a representation of light and heavy nuclei, we take $^{12}$C, $^{56}$Fe
and $^{100}$Mo nuclei and study neutron scattering in the low energy
domain up to 10 MeV. In Fig.\ref{f10} we plot the calculated total cross
sections for all the three systems along with the experimental data 
\cite{rapp,geel,diva,pase}. These results are obtained using TPM15 
parameterization. Each figure has three curves: HOMO (Eq.(\ref{eq13})),
FULL (Eq.(\ref{eq15})) and 1-ITER. The 1-ITER results are shown because,
as found earlier in Section 3, one iteration of Eq.(\ref{eq15}) gives
results very close to the full iteration results. Within the spread in
experimental numbers, all the calculated results are consistent with the
data. In Fig.\ref{f11} we show the various experimental and corresponding
calculated angular distributions. We observe that for $^{12}$C all the
three curves are reasonably consistent with the data. For heavier nuclei,
$^{56}$Fe and $^{100}$Mo, while 1-ITER and FULL results are in good accord
with the data, the HOMO results fall short of it.

\begin{figure*}[htb!]
\centering
\includegraphics[scale=0.36]{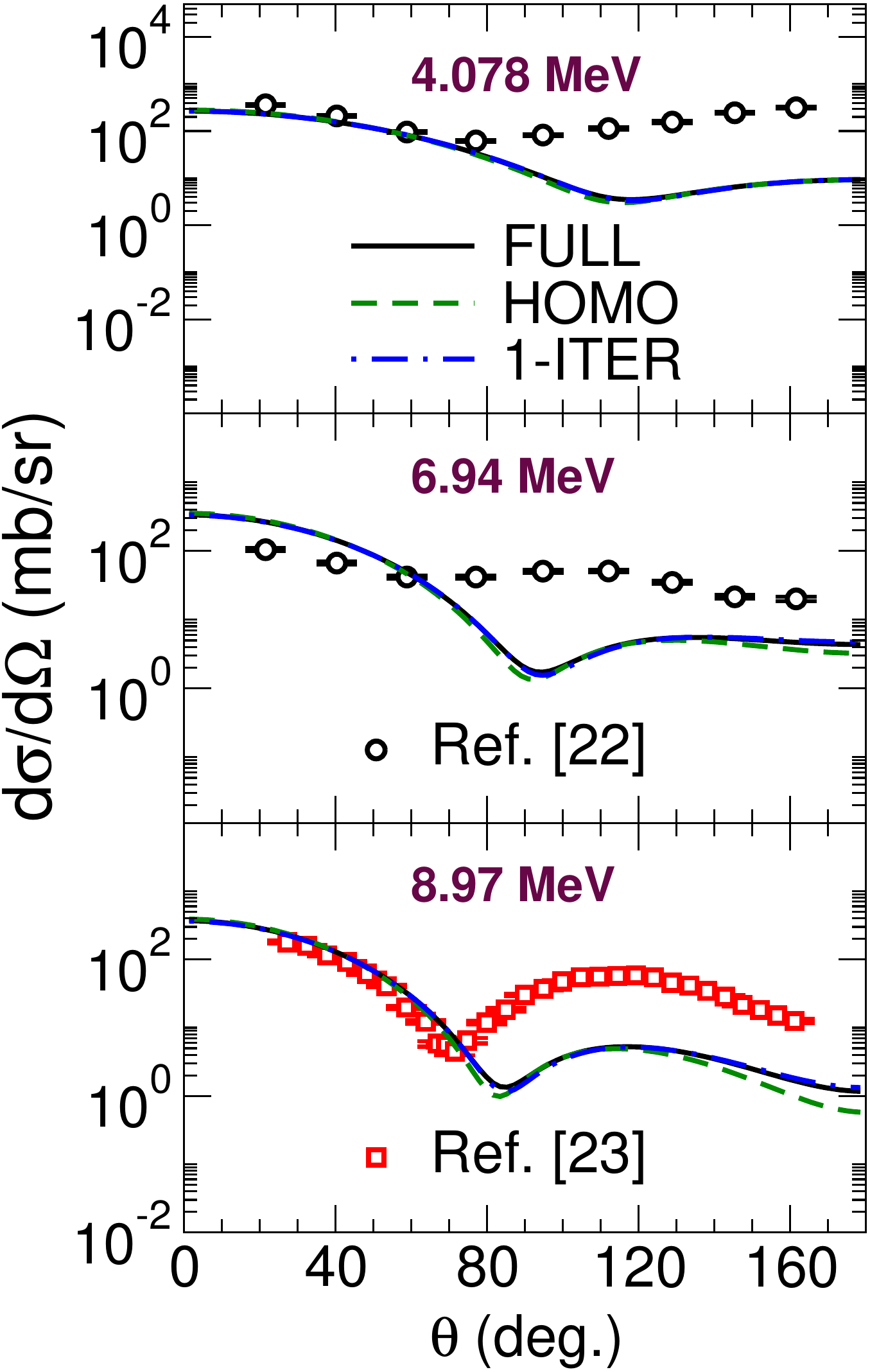}
\caption{Calculated angular distributions for neutron scattering off
$^{12}$C nucleus along with the data \cite{white,glas}. Calculations
are done using FoldTPM15 prescription.}
\label{f12}
\end{figure*}

For $^{12}$C it may be mentioned that, as the TPM15 parameters are obtained
by fitting nucleon scattering data on $^{27}$Al to $^{208}$Pb nuclei, 
probably a better agreement might be achieved if more appropriate choice
of potential is made. To test this possibility, in Fig.\ref{f12} we show 
the calculated angular distribution using FoldTPM15 prescription for 
n-$^{12}$C scattering. These results reproduce the data well only in
the forward hemisphere in magnitude and shape. This indicates that there
is room for improvisation in the mean field sector for such systems. It may
however be mentioned here, as stated earlier in the paper, this improved
agreement may be due to reduced nonlocality displayed in FoldTPM15.

\section{Summary and conclusion}
A novel technique to solve the integro-differential equation in scattering
studies has been developed. It is achieved by applying the mean
value theorem of integral calculus and using the symmetry of the
kernel in the nonhomogeneous term about $r$=$r^\prime$. The extent of
accuracy of the method is established by comparing the solution of the 
homogeneous equation (Eq.(\ref{eq13})) thus obtained, with that of the full
nonhomogeneous equation (Eq.(\ref{eq15})). The later is obtained using
the iterative scheme initiated by solution of the homogeneous equation. It
is found that, though the solution of the homogeneous equation have some
variance with full results, in the first iteration itself this difference
practically disappears. The method is independent of the choice of the 
form of the nonlocality. Hence, it can be used to study the sensitivity to 
the form of the nonlocality in scattering problems. The effective local 
potential appearing in the resultant homogeneous equation is different in 
shape and magnitude in nuclear interior as compared to the local part used 
in the original nonlocal potential. Further, this effective potential is 
found to be $l$-dependent, but energy independent. The total and 
differential cross sections calculated in the low beam energy range (up 
to around 10 MeV) for neutron scattering off $^{12}$C, $^{56}$Fe and 
$^{100}$Mo nuclei compare to a reasonable extent with the corresponding 
measured values. However, the 1-ITER results are found to be in good 
agreement with the data.

\ack
We thank Swagata Sarkar and R. C. Cowsik for a number of illuminating
discussions. We are thankful to the referee for constructive criticism
and several pertinent observations. NJU acknowledges financial support 
from Science and Engineering Research Board (SERB), Govt. of India (grant
number YSS/2015/000900). AB acknowledges financial support from Dept.
of Science and Technology, Govt. of India (grant number 
DST/INT/SWD/VR/P-04/2014).


\end{document}